\documentclass[journal,10pt]{IEEEtran}
\usepackage{amsmath,amsfonts}
\usepackage{algorithmic}
\usepackage{algorithm}
\usepackage{array}
\usepackage{textcomp}
\usepackage{stfloats}
\usepackage{url}
\usepackage{verbatim}
\usepackage{mathrsfs}
\usepackage{amsfonts}
\usepackage{graphicx,cite,epsfig,amssymb,amsmath}
\usepackage{color,xcolor}
\definecolor{red}{rgb}{1.00, 0.00, 0.00}  
\usepackage{pifont}
\usepackage{stmaryrd}
\usepackage{setspace}
\usepackage{subfigure}
\usepackage{cite}
\usepackage{array}
\usepackage{float}
\usepackage{multirow}
\usepackage{epstopdf}
\usepackage{mathtools}
\usepackage{diagbox}
\usepackage{booktabs}
\usepackage{bbm}

\providecommand{\algorithmname}{Algorithm}
\floatname{algorithm}{\protect\algorithmname}
\newcommand{\bm}[1]{\mbox{\boldmath{$#1$}}}

\usepackage{arydshln}
\usepackage{cases}
\usepackage{hyperref}
\usepackage{comment}
\usepackage{setspace}
\begin{document}

\title{Privacy-Preserving Decentralized Inference with Graph Neural Networks in Wireless Networks}
\author{Mengyuan Lee, \IEEEmembership{Student Member, IEEE,} Guanding Yu, \IEEEmembership{Senior Member, IEEE,} and Huaiyu Dai, \IEEEmembership{Fellow, IEEE}
	\thanks{M. Lee  and  G. Yu are with College of Information Science and Electronic Engineering, Zhejiang University, Hangzhou 310027, China. e-mail: \{mengyuan\_lee,  yuguanding\}@zju.edu.cn. (\emph{Corresponding author: Guanding Yu})}
	\thanks{ H. Dai is with the Department of Electrical and Computer Engineering, North Carolina State University,
		Raleigh, NC 27695, USA. e-mail: hdai@ncsu.edu.}
}
\maketitle

\begin{abstract}
As an efficient neural network model for graph data, graph neural networks (GNNs) recently find successful applications for various wireless optimization problems. Given that the inference stage of GNNs can be naturally implemented in a decentralized manner, GNN is a potential enabler for decentralized control/management in the next-generation wireless communications. Privacy leakage, however, may occur due to the information exchanges among neighbors during decentralized inference with GNNs. To deal with this issue, in this paper, we analyze and enhance the privacy of decentralized inference with GNNs in wireless networks. Specifically, we adopt local differential privacy as the metric, and design novel privacy-preserving signals as well as privacy-guaranteed training algorithms to achieve privacy-preserving inference. We also define the SNR-privacy trade-off function to analyze the performance upper bound of decentralized inference with GNNs in wireless networks. To further enhance the communication and computation efficiency, we adopt the over-the-air computation technique and theoretically demonstrate its advantage in privacy preservation. Through extensive simulations on the synthetic graph data, we validate our theoretical analysis, verify the effectiveness of proposed privacy-preserving wireless signaling and privacy-guaranteed training algorithm, and offer some guidance on practical implementation.
\end{abstract}

\begin{IEEEkeywords}
Artificial intelligence, graph neural network, decentralized inference, differential privacy, over-the-air computation.
\end{IEEEkeywords}

\maketitle

\section{Introduction}
With the explosion of mobile users and enhanced QoS requirements, artificial intelligence (AI) techniques gradually become a necessary enabler for the next-generation wireless communications \cite{6g}. In recent years, AI based wireless techniques are attracting increasing interest, and shown to outperform traditional model-based design paradigms in many wireless applications, such as resource allocation \cite{shi,lorm,learntobranch,spatial,gnn_wireless_survey,graphnn,eisen_graph,shen_graph,liu_graph,ca,chen,tmc}, networking \cite{networking1,networking2}, and physical layer design \cite{physical1,physical2,gnn_channel_1,gnn_channel_2,gnn_mimo}. The reasons include the following two aspects. On the one hand, the wireless environments are generally complex due to random channel fading, interference, and hardware impairments. Therefore, the mathematical model sometimes may fail to accurately model the reality of wireless communication systems. On the other hand, large-scale wireless communication schemes, such as the massive multiple-input multiple-output (MIMO) systems, are the mainstream for the next-generation wireless communications.  The corresponding mathematical models inevitably become complex, whose related optimization algorithms exhibit overwhelming computational complexity for practical implementations. AI techniques, however, are data-driven methods with low inference complexity as well as powerful representation ability. Therefore, AI is a potential tool to develop breakthrough wireless techniques for the next-generation wireless communications.

However, different from the success of AI in other fields, such as computer vision and natural language processing, applying AI in wireless networks should meet some special expectations: (i) few labeled training samples since the labeled training samples are difficult to obtain in wireless networks; (ii) high generalization ability for dynamic wireless systems; (iii) feasibility for decentralized implementation in large-scale wireless networks. Amongst existing AI techniques, graph neural networks (GNNs) \cite{gnn_survey1,gnn_survey2} exhibit potential to address the above challenges. First, GNNs can exploit high-dimensional graph topology and node/edge features more efficiently than other neural network models, which enables GNNs to achieve good performance with fewer training samples. Secondly, GNNs can embed each node in the graph to a low-dimensional vector with  functions operated at the node level, which facilitates the same application for graphs of different sizes. Moreover, all node-level operations in GNNs are naturally decentralized, which indicates that the inference stage of well-trained GNN models can be done in the decentralized manner.  Note that most neural network models for wireless applications, such as convolutional neural networks and recurrent neural networks, require central processing during the inference stage. Therefore, the decentralized inference can only be enabled by GNNs. Based on above analysis, GNNs have great potential for efficient and scalable wireless decentralized control.

In this paper, we focus on decentralized inference with GNNs in wireless networks. As mentioned in \cite{tmc}, information exchanges among neighboring nodes are achieved  through wireless channels during decentralized inference with GNNs. Fading and noisy channels may deteriorate the information exchanges and have inevitably adverse impacts on the inference performance. In \cite{tmc}, the authors focus on the possible errors induced by the imperfect wireless transmission and propose a novel  communication-efficient wireless retransmission mechanism to enhance the inference robustness of GNNs, which is essential to bring GNN based wireless techniques from theory to practice. In practical systems, wireless transmission may also incur privacy leakage. Specifically, when neighboring nodes exchange information, personal features may be inferred/detected by honest-but-curious neighbors or possibly anonymous hackers. Generally, personal features may include private information and important parameters that the users do not want to share with others. With the ever-increasing awareness of personal information leakage, privacy preservation for the decentralized inference with GNNs in wireless networks is a significant and non-negligible issue. Therefore, the goal of this paper is to analyze and enhance the privacy of decentralized inference with GNNs in wireless networks.

\subsection{Related Work}
\subsubsection{GNNs in Wireless Networks}
Recently, there are some existing works about incorporating GNNs into wireless applications \cite{gnn_wireless_survey,graphnn,eisen_graph,shen_graph,liu_graph,ca,chen,gnn_channel_1,gnn_channel_2,gnn_mimo}. Most of them have employed GNNs for  wireless resource allocation, such as link scheduling \cite{graphnn}, power allocation \cite{eisen_graph,shen_graph} and user association \cite{liu_graph}. Some have paid attention to physical layer design, such as channel estimation \cite{gnn_channel_1,gnn_channel_2} and MIMO detection \cite{gnn_mimo}. In these existing works, GNN based solutions are generally implemented in the centralized server. Meanwhile, other existing works in literature have used GNNs for decentralized control, such as multi-robot management \cite{control1,control2,control3} and wireless channel allocation \cite{control4}. However, they generally assume perfect information exchanges between nodes and do not pay attention to possible privacy leakage issue.

\subsubsection{Privacy Preservation for GNNs}
Existing techniques targeting privacy preservation of GNNs can be summarized as follows: (i) making use of collaborative learning schemes, such as  federated/split learning, to protect the training privacy of GNNs \cite{fedprivacy,splitprivacy}; (ii) modifying existing training algorithms, such as adding noise to the gradients generated by the SGD/Adam algorithm, for preserving the training privacy of GNNs \cite{private_adam,private_sgd,private_noise}; (iii) adding noise or coding on the private information, such as node/edge features, to achieve privacy-preserving GNN's inference \cite{private_coding}.  Compared to privacy protection of GNNs at the training stage, much less is done at the inference stage, and none of existing works takes wireless environments into consideration. This motivates us to study the role of wireless fading and noise for privacy preservation and develop novel privacy-preserving techniques for decentralized inference with GNNs in wireless networks.

\subsubsection{Privacy Preservation in Wireless Networks}
Some existing works in literature have already paid attention to privacy preservation by exploiting wireless fading and noise \cite{wirelessprivate1,wirelessprivate2,wirelessprivate3,wirelessprivate4,wirelessprivate5,wirelessprivate6,wirelessprivate7}. However, all of them focus on the federated learning scenario and protect the privacy for distributed/decentralized training, rather than the inference stage. Therefore, these aforementioned existing methods are only applicable to distributed/decentralized training of GNNs in wireless networks. Note that the characteristics of the training and inference stages are  different. First, the training stage makes use of stochastic algorithms (such as SGD) and is inherently tolerant of noise. Inference stage, however, is a deterministic process and more vulnerable to noise. Therefore, exploiting noise for privacy preservation imposes a more severe impact on the inference stage than on the training stage. Second, the performance metric of the training stage is the well-studied ``convergence rate" while that of the inference stage is difficult to be expressed in closed form. To some extent, privacy preservation at the inference stage with performance guarantee incurs more challenges and requires further considerations compared with that at the training stage. Therefore, these aforementioned works are not applicable to decentralized inference with GNNs in wireless networks, which motivates the study in this paper.

\subsection{Contributions and Outline}
As far as we know, this is the first piece of work to study privacy-preserving decentralized inference with GNNs in wireless networks. Specifically, our main contributions are summarized as follows.

{\bf 1. Privacy-Preserving Wireless Signaling:} We adopt local differential privacy as the privacy-preserving metric and propose novel wireless signaling to achieve privacy-preserving decentralized inference with GNNs in wireless networks. Based on the optimal parameter solutions for privacy-preserving signaling, we define the SNR-privacy trade-off function to theoretically analyze the performance upper bound of the proposed wireless signaling.

{\bf 2. Privacy-Guaranteed Training Algorithm:} Traditional training algorithms may incur inference performance loss when adopting the proposed privacy-preserving wireless signaling. Therefore, we design privacy-guaranteed training algorithm to further enhance the performance of privacy-preserving decentralized inference with GNNs in wireless networks.

{\bf 3. Advantages of AirComp on Privacy Preservation:} We exploit the over-the-air computation (Aircomp) technique for the decentralized inference with GNNs in wireless networks, and theoretically and experimentally demonstrate that it can enhance the communication and computation efficiency as well as the privacy preservation.

The rest of this paper is organized as follows. In Section \ref{section:pre}, we introduce preliminaries and backgrounds for some important techniques we adopted in this paper. In Section \ref{section:system}, we formulate the wireless optimization problem, introduce its corresponding graphical model, and discuss the decentralized implementation of GNNs to solve it. In Section \ref{section:signal}, privacy-preserving wireless signaling is proposed, based on which the SNR-privacy trade-off is analyzed. The privacy-guaranteed training algorithm is introduced in Section \ref{section:training}.
Extension analysis to the system without the Aircomp technique and advantages of the Aircomp technique are discussed in Section \ref{section:aircomp}. 
The simulation results are given in Section \ref{section:simulation}. Finally, we conclude the paper and provide directions for future research in Section \ref{section:con}. For ease of reading, the notations used in this paper are summarized in Table \ref{table:notations}.

\begin{table*}
	\vspace{-2em}
	\footnotesize
	\caption{Notations in this Paper}
	\vspace{-1em}
	\label{table:notations}
	\centering
	\begin{tabular}{|c|c|}
		\hline
		Notations & Meaning \\
		\hline
		$G$ & Graph \\
		\hline
		$\mathcal{V}$ & Node set \\
		\hline
		$\mathcal{E}$ & Edge set \\
		\hline
            $N(v)$& Neighbors of node $v$\\
            \hline
		$\bm{x}_v$ & Features of node $v$ \\
		\hline
		$\bm{e}_{vu}$ & Features of edge $e(v,u)$ \\
		\hline
		$f_M^{(k)}(\cdot)$ & Local  message function at the $k$-th layer \\
		\hline
            $f_U^{(k)}(\cdot)$ & Local update function at the $k$-th layer \\
		\hline
            ${\rm Agg}(\cdot)$ & Aggregation function \\
            \hline
            $\bm{h}_v^{(k)}$/$\bm{\hat{h}}_v^{(k)}$ & Hidden state/Message of node $v$ at the $k$-th layer\\
            \hline
            $g_{vu}$& Channel gain from node $v$ to $u$\\
            \hline
            $P_v$ &  Maximum transmit power of node $v$\\
            \hline
            $\bm{w}_{uv}^{(k)}$/$\bm{\tilde{w}}_{uv}^{(k)}$ & Original/Privacy-preserving signal from node $u$ to $v$ at the $k$-th iteration\\
            \hline
            $\bm{R}_v^{(k)}$/$\bm{\tilde{R}}_v^{(k)}$& Received signal of node $v$ at the $k$-th iteration with original/privacy-preserving signal\\
            \hline
            $\gamma_{uv}$, $\alpha_{uv}$,  $\beta_{uv}$ & Parameters for privacy-preserving signals from node $u$ to $v$
            \\
            \hline
            $\bm{m}_{uv}$& Artificial Gaussian noise vector for privacy-preserving signals from node $u$ to $v$\\
            \hline
            $C_v$& Aligned amplitudes for signals transmitted from the neighbors of node $v$\\
            \hline
            $\epsilon_v^*$& Pre-given target privacy budget \\
            \hline
            $\rho_v$ & SNR of $\bm{\tilde{R}}_v^{(1)}$\\
            \hline
	\end{tabular}
 \vspace{-1em}
\end{table*}

\section{Preliminaries} \label{section:pre}
\subsection{Graph Neural Networks} \label{section:gnn_pre}
\begin{figure*}
\centering	
\includegraphics[width=0.7\linewidth]{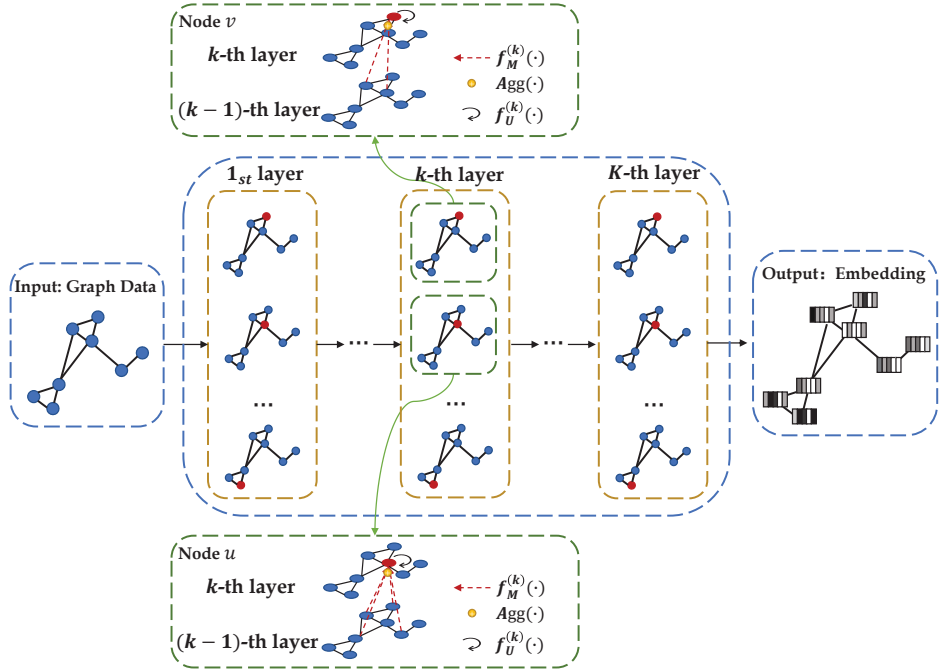}
\vspace{-1em}
\caption{Graph illustration for the structure of GNNs.}
\label{fig:gnn}
\vspace{-1em}
\end{figure*}

GNNs, as the name implies, are especially designed for graph data over the non-Euclidean domain \cite{gnn_survey1,gnn_survey2}. By taking the graph adjacency matrix and the node/edge features as the input, GNNs follow the layer-wise structure and output embedding vectors for each node in the graph. The graphical illustration of GNNs can be found in Fig. \ref{fig:gnn}. Specifically, each layer of GNNs consists of the following two steps: (i) local  message function and aggregation function that collect information from the neighborhood of each node in the graph; (ii) local update function that updates the hidden state of each node in the graph at a certain layer. Note that the local  message and update functions are the same for all nodes in the graph and are generally realized through multi-layer perceptrons (MLPs). For a graph $G(\mathcal{V},\mathcal{E})$ with node features $\{\bm{x}_v\}_{v\in\mathcal{V}}$ and edge features $\{\bm{e}_{vu}\}_{e(v,u)\in\mathcal{E}}$, the operations at the $k$-th layer of GNNs can be mathematically presented as follows
\begin{equation}
	\bm{h}_v^{(k)} = f_U^{(k)}(\bm{h}_v^{(k-1)},{\rm Agg}_{u\in N(v)}(f_M^{(k)}(\bm{h}_u^{(k-1)},\bm{e}_{vu}))),\label{local_out} 
\end{equation}
where $f_M^{(k)}(\cdot)$ and $f_U^{(k)}(\cdot)$ are the local message and update functions at the $k$-th layer, respectively\footnote{$M$ and $U$ are initials of ``message" and ``update", respectively.}. ${\rm Agg}(\cdot)$ is the aggregation function whose popular candidates include $\sum(\cdot)$, ${\rm mean}(\cdot)$, and $\max(\cdot)$. $\bm{h}_v^{(k)}$ represents the hidden state of node $v$ at the $k$-th layer. Note that $\bm{h}_v^{(0)}=\bm{x}_v$. Moreover, $N(v)$ denotes the neighbors of node $v$. By stacking multiple layers, the local information of each node in the graph propagates through the network. The final embedding vectors $\{\bm{h}_v^{(K)}\}_{v\in \mathcal{V}}$ can be used for further node/edge/graph-level analysis.

\subsection{Local Differential Privacy} \label{section:ldp_pre}
Local differential privacy (LDP) generally serves as a strong criterion of preserving privacy in decentralized systems, which finds successful applications in many commercial companies, such as Google \cite{google} and Microsoft \cite{Microsoft}.  The key idea of LDP is enabling statistical analysis without divulging the personal information, which is more than data security. Specifically, LDP can be defined formally as follows.

{\bf{Definition 1:}} ($(\epsilon, \delta)$-LDP \cite{dp}) A randomized mechanism $\mathcal{M}:\mathbb{N}^{|\mathcal{X}|}\rightarrow \mathbb{R}^d$ is $(\epsilon, \delta)$-LDP if for any two adjacent databases $\bm{x}, \bm{x}'\in \mathbb{N}^{|\mathcal{X}|}$ and any measurable subset $\mathcal{O} \subseteq  Range(\mathcal{M})$, we have
$Pr(\mathcal{M}(\bm{x})\in \mathcal{O}) \leq e^\epsilon Pr(\mathcal{M}(\bm{x}')\in \mathcal{O})+\delta,$
where $\epsilon$ is called privacy budget and controls the similarities between the probabilities that the mechanism $\mathcal{M}$ assigns to the outputs of adjacent databases $\bm{x}, \bm{x}'$. The smaller the $\epsilon$ is, the more difficult an adversary could distinguish two adjacent databases with high probability by only looking at the output. Therefore, a smaller  $\epsilon$ leads to stronger  privacy guarantee. Moreover, $\delta$ denotes the probability of violating the above privacy guarantee. Similarly,  a smaller   $\delta$ also leads to stronger privacy guarantee. 

According to how adjacent databases are defined, LDP can be categorized into client-level LDP and record-level LDP that provide privacy preservation for one client's records and a single record of one client, respectively \cite{client_dp}. Generally, client-level LDP provides stronger privacy preservation than record-level LDP. Therefore, we use client-level LDP in this paper to protect node features for decentralized inference with GNNs in wireless networks and define that two graphs are neighboring if only one node feature differs.

Furthermore, for numerical data, $(\epsilon, \delta)$-LDP can be achieved by using the Gaussian mechanism defined as follows.

{\bf{Definition 2:}} (Gaussian Mechanism \cite{dp}) Let $f(\cdot):\mathbb{N}^{|\mathcal{X}|}\rightarrow \mathbb{R}^d$ be an arbitrary function and define its $l_2$ sensitivity to be bounded by $\Delta_{2f}$, i.e., $||f(\bm{x})-f(\bm{x}')||_2 \leq \Delta_{2f}, \forall \text{adjacent } \bm{x}, \bm{x}'\in \mathbb{N}^{|\mathcal{X}|}$. The Gaussian mechanism is defined as
$M(\bm{x})\triangleq f(\bm{x})+\mathcal{N}(0,\sigma^2\bm{I}),$
where $\mathcal{N}(0,\sigma^2\bm{I})$ denotes the $d$-dimensional Gaussian noise whose mean is 0 and variance is $\sigma^2$. 

Moreover, for any $\epsilon \in (0,1]$ and $\delta \in (0,1]$, the above defined Gaussian mechanism is $(\epsilon, \delta)$-LDP, where
$\epsilon=\Delta_{2f}\sqrt{2\ln \frac{1.25}{\delta}}/\sigma$.
Given that noise generally follows Gaussian distribution in communication systems, we will make use of the Gaussian mechanism in this paper to achieve privacy-preserving decentralized inference with GNNs in wireless networks.

\subsection{Over-the-Air Computation}
Over-the-air computation (Aircomp) recently becomes prevalent in distributed learning schemes, such as  federated learning  \cite{FL_aircomp,FL_aircomp2}. By exploiting the signal-superposition property, Aircomp can simultaneously transmit and compute the analog signals from users in a multiple-access channel (MAC), and thus significantly enhance the
communication and computation efficiency compared with the orthogonal transmit-then-compute protocol. Using an ideal MAC with $K$ users as an example, each user transmits a signal $s_k\in \mathbb{R}$, and the received signal at the receiver with signal-superposition property is given by $r= \sum_{k=1}^{K} s_k,$ which achieves the task of computing the sum of $K$ users' signals. Similarly, by additionally adopting pre- and post-processing functions, the above Aircomp process can also achieve other real-valued multivariate functions, such as the arithmetic/geometric mean.

The key observation for GNNs is that aggregating neighborhood information to update the hidden state of each node in the graph falls in the category of computing functions of distributed data if ${\rm Agg}(\cdot)$ is set as  $\sum(\cdot)$ or ${\rm mean}(\cdot)$. Therefore, in this paper, we propose adopting the Aircomp technique for communication- and computation-efficient aggregation during the inference stage of GNNs.

\section{System Model} \label{section:system}
In this section, we first introduce the general formulation of the wireless optimization problem and its corresponding graphical model. Then we discuss using GNNs for the wireless optimization problem as well as its decentralized implementation in wireless networks.

\subsection{Wireless Optimization Problem and Corresponding Graphical Model}
\begin{figure}
	\vspace{-1em}
	\centering
\includegraphics[width=1\linewidth, height=0.14\textheight]{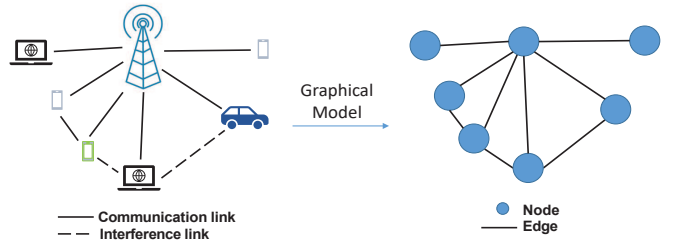}
	\vspace{-1em}
	\caption{A graphical illustration of the wireless network (left) and its graphical model (right).}
	\label{fig:model}
	\vspace{-2em}
\end{figure}

Wireless optimization problems can be generally formulated as optimizing over graph data, $G(\mathcal{V},\mathcal{E})$, with the node set $\mathcal{V}$ and the edge set $\mathcal{E}$\footnote{$G(\mathcal{V},\mathcal{E})$ can be either directed or undirected according to specific wireless optimization problems. Moreover, if $G(\mathcal{V},\mathcal{E})$ is dynamic but remains static within the coherence time, our proposed method is applicable for each instant of the dynamic graph.}. According to specific applications, nodes in $\mathcal{V}$ correspond to wireless devices, such as mobile devices, edge servers, access points, or base stations, while edges in $\mathcal{E}$ indicate that two nodes (devices) can communicate or interfere with each other. Moreover, all nodes and edges may have corresponding features, denoted as $\{\bm{x}_v\}_{v\in \mathcal{V}}$ and $\{\bm{e}_{vu}\}_{e(v,u)\in \mathcal{E}}$, respectively. Specifically, the features for nodes and edges include the needed characteristics/hyperparameters for wireless optimization problems, such as the power limit of each user and the channel gain between two users. The node features may also include extra sensitive and private information such as gender and self-preference that the wireless devices do not want to share with others. With the above graphical model, wireless optimization problems can be  formulated as node/edge/graph-level analysis tasks. For example, link scheduling \cite{graphnn} corresponds to node-level classification and power control \cite{eisen_graph,shen_graph} corresponds to node-level regression. A graphical illustration of the above process can be found in Fig. \ref{fig:model}.

\subsection{Decentralized Implementation of GNNs for Wireless Optimization Problems}
With the graphical model, we can use the $K$-layer GNN introduced in Section \ref{section:gnn_pre} to solve  the wireless optimization problem. Specifically, with a series of operations in  (\ref{local_out}), each node in $\mathcal{V}$ is embedded as a vector. With additional MLPs or other learning models (e.g. reinforcement learning) to further process the embedding vectors of all nodes, corresponding node/edge/graph-level analysis tasks can be accomplished. Without loss of generality, we adopt $\sum(\cdot)$ as the aggregation function for the subsequent discussion of this paper, then the operations of the $k$-th layer of the GNN are given by
\begin{equation}
	\bm{\hat{h}}_v^{(k)} = \sum_{u\in N(v)}f_M^{(k)}(\bm{h}_u^{(k-1)},\bm{e}_{vu}), \label{sum_tran}
\end{equation}
\begin{equation}
	\bm{h}_v^{(k)} = f_U^{(k)}(\bm{h}_v^{(k-1)},\bm{\hat{h}}_v^{(k)}),\label{sum_out} 
\end{equation}
where $\bm{\hat{h}}_v^{(k)}$ is termed the message of node $v$ at the $k$-th laye.

Note that the operations in (\ref{sum_tran}) and (\ref{sum_out}) are naturally decentralized. Meanwhile,  deploying the inference stage of GNNs in a decentralized manner has been actively researched  in literature \cite{control1,control2,control3}. As a premise, we assume that the GNN model is already well-trained and each node (user) in the graph keeps a copy of it, which means that the
parameters of the GNN model are available at each node\footnote{The detailed privacy-guaranteed training algorithm for GNNs will be discussed in Section \ref{section:training}}. During the inference stage, each node $v\in \mathcal{V}$ can accomplish its own prediction and solve the corresponding wireless optimization problem in a decentralized manner by repeating the following process for $K$ times. Specifically, for the $k$-th iteration, 
\begin{itemize}
	\item Node $v$ sends transmission requests to each node $u \in N(v)$.
	\item After receiving the request from node $v$, each node $u \in N(v)$ transmits a signal 
	$\bm{w}_{uv}^{(k)}=\sqrt{P_u}f_M^{(k)}(\bm{h}_u^{(k-1)},\bm{e}_{vu})$
	to node $v$ through wireless channels, where $P_u$ is the maximum transmit power of node $u$. 
	\item Node $v$ computes the sum of $\{f_M^{(k)}(\bm{h}_u^{(k-1)},\bm{e}_{vu})\}_{u \in N(v)}$ to get $\bm{\hat{h}}_v^{(k)}$, and then imposes $ f_U^{(k)}(\cdot)$ to get $\bm{h}_v^{(k)}$.
\end{itemize}
Finally, the embedding vector for node $v$, i.e., $\bm{h}_v^{(K)}$, can be obtained for further analysis.

\subsection{Wireless Communication Model} \label{section:communication_model}
The information exchanges among neighbors are achieved through wireless channels. In this paper, we consider a single-antenna multiple-access system with slow fading and independent and identically distributed (i.i.d.)  Gaussian noise, which indicates that the channel gain remains static within the inference while the noise is i.i.d over different iterations. We further assume that the channel state information (CSI) is available to both the transmitter and receiver. With the Aircomp technique, the sum of $\{f_M^{(k)}(\bm{h}_u^{(k-1)},\bm{e}_{vu})\}_{u\in N(v)}$ is computed during the communication process. Therefore, the received signal of node $v$ at the $k$-th iteration is given by
\begin{equation*}
	\begin{aligned}
		\bm{R}_v^{(k)} &= \sum_{u\in N(v)} g_{uv}\bm{w}_{uv}^{(k)}+\bm{n}_v^{(k)},\\&= \sum_{u\in N(v)} g_{uv}\sqrt{P_u}f_M^{(k)}(\bm{h}_u^{(k-1)},\bm{e}_{vu})+\bm{n}_v^{(k)},
	\end{aligned}
\end{equation*}
where $g_{uv}=|g_{uv}|e^{j\varphi_{uv}}$ is the complex valued channel coefficient from node $u$ to node $v$, and $\bm{n}_v^{(k)}$ is the Gaussian noise vector whose elements are i.i.d drawn from $\mathcal{N}(0,\sigma_v^2)$ for node $v$ at the $k$-th iteration\footnote{Following existing literature about the Aircomp, neither interference nor channel coding is considered in this work so as to focus on the main theme of privacy protection in GNN inference. The proposed mechanism can be naturally extended to scenarios involving interference and channel coding with some minor modifications.}.

The above transmission scheme exhibits the following two issues. First, the received signal $\bm{R}_v^{(k)}$ is contaminated by fading and noise, and therefore cannot be directly used to estimate $\bm{\hat{h}}_v^{(k)}$ in (\ref{sum_tran}). Moreover, the personal features of each node, i.e. $\bm{x}_u$(or say $\bm{h}_u^{(0)})$, may be inferred/detected by honest-but-curious neighbors and possibly anonymous hackers from the transmitting signal $\bm{w}_{uv}^{(1)}$. In the following, we propose novel techniques, including privacy-preserving wireless signaling and privacy-guaranteed training algorithm, to overcome the above two challenges to achieve  privacy-preserving decentralized inference with GNNs in wireless networks.

\section{Privacy-Preserving Wireless Signaling} \label{section:signal}
In this section, we introduce the privacy-preserving wireless signaling  aiming at  privacy-preserving decentralized inference with GNNs in wireless networks. We first formulate the privacy-preserving wireless signaling problem and then derive its optimal solution, based on which the SNR-privacy trade-off is analyzed to provide performance upper bound. 
\subsection{Problem Formulation} \label{section:problem_formualation}
\subsubsection{The Last $K-1$ Iterations} \label{new_section}
Our focus is to protect the privacy of node features. For the last $K-1$ iterations during the inference, node features are not directly involved in the computation of exchanged information, i.e. $\{\bm{w}_{uv}^{(k)}\}_{k=2}^K$. Therefore, node features are difficult to be inferred if the information of the first iteration is missing. Based on this fact, no privacy leakage exists for the last $K-1$ iterations\footnote{If one still wants to enhance the privacy for the last $K-1$ iterations, the proposed privacy-preserving wireless signaling can still help with optimizing the signal in a similar way as the first iteration.} and the corresponding goal of signal design is to correct channel fading and noise to get unbiased and accurate estimations on $\{\bm{\hat{h}}_v^{(k)}\}_{k=2}^K$. Thus, we propose the following signal for node $u$ communicating with node $v$ at the $k$-th iteration
\begin{equation}
\bm{\tilde{w}}_{uv}^{(k)}=e^{-j\varphi_{uv}}\sqrt{\gamma_{uv}P_u}f_M^{(k)}(\bm{h}_u^{(k-1)},\bm{e}_{vu}),  \label{each_user_1}
\end{equation}
where the multiplication factor $e^{-j\varphi_{uv}}$ is used to correct the phase shift induced by the wireless channel, while $\gamma_{uv}\in(0,1]$ is a parameter to be optimized. With this newly designed signal, the received signal of node $v$ at the $k$-th iteration is given by 
\begin{equation}
	\begin{aligned}
		\bm{\tilde{R}}_v^{(k)} &= \sum_{u\in N(v)} g_{uv}\bm{\tilde{w}}_{uv}^{(k)}+\bm{n}_v^{(k)},\\&= \sum_{u\in N(v)} |g_{uv}|\sqrt{\gamma_{uv}P_u}f_M^{(k)}(\bm{h}_u^{(k-1)},\bm{e}_{vu})+\bm{n}_v^{(k)}.
	\end{aligned}
	\label{rv1}
\end{equation}
By following the rules of Aircomp, the amplitudes of all neighbors' information need to be aligned as a constant to obtain an unbiased estimation on $\bm{\hat{h}}_v^{(k)}$, i.e., $|g_{uv}|\sqrt{\gamma_{uv}P_u}=\text{constant}, \forall u\in N(v)$. On the other hand, the aligned constant needs to be maximized to enhance the estimation accuracy. Therefore, the optimal solution for $\gamma_{uv}$ is given by
\begin{equation}\gamma_{uv}=\frac{\min_{u\in N(v)}|g_{uv}|^2P_u}{|g_{uv}|^2P_u}. \label{final_noprivacy}\end{equation}

\subsubsection{The First Iteration}
For the first iteration, the proposed wireless signaling is supposed to not only enable the  estimation on $\bm{\hat{h}}_v^{(1)}$ but also provide privacy guarantee on the features of neighbors, i.e. $\{\bm{x}_u\}_{u\in N(v)}$. As introduced in Section \ref{section:ldp_pre}, Gaussian mechanism is efficient for numerical data to achieve $(\epsilon, \delta)$-LDP. Therefore, by adding Gaussian noise to the raw transmitting signal $\bm{w}_{uv}^{(1)}$, the privacy-preserving transmitting signal from node $u$ to node $v$ at the first iteration is proposed as
\begin{equation}
	\bm{\tilde{w}}_{uv}^{(1)}=e^{-j\varphi_{uv}}[\sqrt{\alpha_{uv}P_u}\check{f}_M^{(1)}(\bm{h}_u^{(0)},\bm{e}_{vu})+\sqrt{\beta_{uv}P_u}\bm{m}_{uv}], \label{each_user_signal}
\end{equation}
where $\bm{m}_{uv}$ is the artificial Gaussian noise vector for privacy preservation whose elements are i.i.d and follow the normal distribution, $\mathcal{N}(0,1)$. $\check{f}_M^{(1)}(\bm{h}_u^{(0)},\bm{e}_{vu})$ is the normalized version of $f_M^{(1)}(\bm{h}_u^{(0)},\bm{e}_{vu})$, i.e., $||\check{f}_M^{(1)}(\bm{h}_u^{(0)},\bm{e}_{vu})||_2=1$. Moreover, $\alpha_{uv}$ and $\beta_{uv}$ denote the fractions of power assigned to the information of node $u$ and the artificial Gaussian noise, respectively, which satisfies  $0<\alpha_{uv}\leq 1$ and $0 \leq \beta_{uv}\leq 1-\alpha_{uv}$\footnote{Note that different from existing signaling, $\alpha_{uv}+\beta_{uv}=1$ does not always hold in our considered problem, which will be further validated by the following analysis. \label{assumption}}. By adopting the newly designed signal, the received signal of node $v$ at the first iteration is given by
\begin{equation}
	\begin{aligned}
		\bm{\tilde{R}}_v^{(1)} = &\sum_{u\in N(v)} |g_{uv}|\sqrt{\alpha_{uv}P_u}\check{f}_M^{(1)}(\bm{h}_u^{(0)},\bm{e}_{vu})\\&+
		\sum_{u\in N(v)} |g_{uv}|\sqrt{\beta_{uv}P_u}\bm{m}_{uv}+\bm{n}_v^{(1)},
	\end{aligned}
\label{rv}
\end{equation}
where the first part is used for estimating  $\bm{\hat{h}}_v^{(1)}$ while the last two noisy parts are used for privacy preservation.

{\bf Goal 1-Estimation:} Similar to the analysis in Section \ref{new_section},  the amplitudes of all neighbors' information need to be aligned as a constant $C_v$ to obtain the estimation on $\bm{\hat{h}}_v^{(1)}$, i.e.,
\begin{equation}
|g_{uv}|\sqrt{\alpha_{uv}P_u}=C_v, \forall u\in N(v).\label{alpha-c}
\end{equation}
On the other hand, to enhance the accuracy of the estimation on $\bm{\hat{h}}_v^{(1)}$, the SNR of $\bm{\tilde{R}}_v^{(1)}$, defined as the ratio between the power of  efficient signal and noise, needs to be maximized. Specifically, the SNR of $\bm{\tilde{R}}_v^{(1)}$, denoted as $\rho_v$, is mathematically given by  
\begin{equation}
	\rho_v=\frac{C_v^2}{\sum_{u\in N(v)}|g_{uv}|^2\beta_{uv}P_u+\sigma_v^2}.
\label{SNR}
\end{equation}

{\bf Goal 2-Privacy preservation:} We introduce the following theorem for analyzing the privacy guarantee   achieved by $\bm{\tilde{R}}_v^{(1)}$ for each node in $N(v)$, which is proved in Appendix \ref{epsion1}.

{\bf{Theorem 1:}} For each node $u$ in $N(v)$, $\bm{\tilde{R}}_v^{(1)}$ achieves $(\epsilon_v, \delta)$-LDP when predicting node $v$'s result, where\begin{small}$$\epsilon_v=\frac{2C_v}{\sqrt{\sum_{u\in N(v)}|g_{uv}|^2\beta_{uv}P_u+\sigma_v^2}}\sqrt{2\ln \frac{1.25}{\delta}}. $$\end{small}
To achieve privacy-preserving inference, the privacy guarantee $\epsilon_v$ needs to be smaller than a pre-given target budget $\epsilon_v^*$.

Based on above analysis, the signal design problem for the first iteration can be formulated as follows
\begin{equation}\label{Design_problem}
	\rho_v^{\max} \triangleq \max_{\{\alpha_{uv},\beta_{uv}\}_{u\in N(v)}} \rho_v=\frac{C_v^2}{\sum_{u\in N(v)}|g_{uv}|^2\beta_{uv}P_u+\sigma_v^2}, 
\end{equation}
subject to (\ref{alpha-c}) and
\begin{align}
	0<\alpha_{uv}\leq 1, 0 \leq \beta_{uv}\leq 1-\alpha_{uv}, \forall u\in N(v), \label{cons1}\tag{\theequation a}
\end{align}
\vspace{-2em}
\begin{align}
\frac{2C_v}{\sqrt{\sum_{u\in N(v)}|g_{uv}|^2\beta_{uv}P_u+\sigma_v^2}}\sqrt{2\ln \frac{1.25}{\delta}} \leq \epsilon_v^*.
\label{privacy_constraints}\tag{\theequation b}
\end{align}

\subsection{Optimal Solution for Problem (\ref{Design_problem})} \label{section:optimal_solution}
In this subsection, we will develop the optimal solution for Problem (\ref{Design_problem}). First, we will make use of the aligned amplitude to transform Problem (\ref{Design_problem}) into a more tractable form. Then, we will discuss and derive the optimal solution when $\epsilon_v^*$ is in different ranges.
 
 Specifically, from (\ref{alpha-c}), there exists one-to-one correspondence between the values of $\{\alpha_{uv}\}_{u\in N(v)}$ and $C_v$.
Therefore, (\ref{cons1}) can be rewritten as

\begin{small}
\begin{equation}
	\begin{aligned}
     \left\{
		\begin{array}{lr}
			0< C_v \leq \min_{u\in N(v)}\sqrt{|g_{uv}|^2P_u},\\ 0 \leq \beta_{uv}\leq 1-\frac{C_v^2}{|g_{uv}|^2P_u}, \forall u\in N(v).
		\end{array}
		\right.
	\end{aligned}
\label{new_cons}
\end{equation}
\end{small}With the above analysis, Problem (\ref{Design_problem}) can be rewritten as
\begin{equation}\label{Design_problem2}
	\rho_v^{\max} \triangleq \max_{C_v, \{\beta_{uv}\}_{u\in N(v)}} \rho_v=\frac{C_v^2}{\sum_{u\in N(v)}|g_{uv}|^2\beta_{uv}P_u+\sigma_v^2}, 
\end{equation}
subject to (\ref{privacy_constraints}) and (\ref{new_cons}).

By looking into (\ref{new_cons}), we have
\begin{equation*}
	\begin{aligned}
		\rho_v&\leq \frac{(\min_{u\in N(v)}(\sqrt{|g_{uv}|^2P_u})^2}{\sum_{u\in N(v)}|g_{uv}|^2\times0\times P_u+\sigma_v^2}
		=\frac{\min_{u\in N(v)}|g_{uv}|^2P_u}{\sigma_v^2}.
	\end{aligned}
\end{equation*} 
On the other hand, with (\ref{privacy_constraints}), we have $\rho_v \leq {\epsilon_v^*}^2/(8\ln \frac{1.25}{\delta})$. Therefore,
$$\rho_v^{\max}=\min\{\frac{\min_{u\in N(v)}|g_{uv}|^2P_u}{\sigma_v^2},\frac{{\epsilon_v^*}^2}{8\ln \frac{1.25}{\delta}}\}.$$
The two upper bounds coincide with each other when
$$\epsilon_v^*=\sqrt{\frac{8\ln \frac{1.25}{\delta}\min_{u\in N(v)}|g_{uv}|^2P_u}{\sigma_v^2}}\triangleq \epsilon_v^{(0)}.$$ 
In the following, we discuss the solution to Problem (\ref{Design_problem2}) under two different cases.

\subsubsection{$\epsilon_v^* > \epsilon_v^{(0)}$}
When $\epsilon_v^* > \epsilon_v^{(0)}$ holds, we have   $\rho_v^{\max} =\min_{u\in N(v)}|g_{uv}|^2P_u/\sigma_v^2$ with
$C_v= \min_{u\in N(v)}\sqrt{|g_{uv}|^2P_u},$ and $\beta_{uv}=0,  \forall u\in N(v).$			
The solution indicates that no artificial noise is needed and the target privacy guarantee can be satisfied by only exploiting the channel noise\footnote{In case 1, $\alpha_{uv}$ depends on  the worst-case received power from neighbors, i.e, $\min_{u\in N(v)}|g_{uv}|^2P_u$. Meanwhile $\beta_{uv}=0$ because the privacy-preserving requirement is weak. With the above solutions, $\alpha_{uv}+\beta_{uv}=1$ does not always hold for all neighbors in $N(v)$, which validates the statement in Section \ref{section:problem_formualation}.}.

\subsubsection{$\epsilon_v^* \leq \epsilon_v^{(0)}$}
When $\epsilon_v^* \leq \epsilon_v^{(0)}$ holds, we have  $\rho_v^{\max} ={\epsilon_v^*}^2/(8\ln \frac{1.25}{\delta})$. The feasible solutions of $C_v$ and $\{\beta_{uv}\}_{u\in N(v)}$ fall into the set given by (\ref{rset1}) - (\ref{rset3}),
\begin{figure*}[tb]
\small
\begin{subnumcases}{\Phi_v=}
	0< C_v \leq \min\{\min_{u\in N(v)}\sqrt{|g_{uv}|^2P_u}, \epsilon_v^*\sqrt{\frac{\sigma_v^2+\sum_{u\in N(v)}|g_{uv}|^2P_u}{8\ln \frac{1.25}{\delta}+|N(v)|{\epsilon_v^*}^2}}\}, & \label{rset1} \\
	0 \leq \beta_{uv}\leq 1-\frac{C_v^2}{|g_{uv}|^2P_u}, \forall u\in N(v), &\label{rset2} \\
	\frac{2C_v}{\sqrt{\sum_{u\in N(v)}|g_{uv}|^2\beta_{uv}P_u+\sigma_v^2}}\sqrt{2\ln \frac{1.25}{\delta}} = \epsilon_v^*,&\label{rset3}
\end{subnumcases}
\hrulefill
\end{figure*}
whose mathematical derivation can be found in Appendix \ref{appendix:phi}. When adopting the Aircomp technique, the aligned amplitude, i.e., $C_v$, is preferred to be as large as possible \cite{wirelessprivate1}. Therefore, the solution of $C_v$ is given by 

\begin{small}
\begin{equation}
	\begin{aligned}
		C_v 
		&= \left\{
		\begin{array}{lr}
			\epsilon_v^*\sqrt{\frac{\sigma_v^2+\sum_{u\in N(v)}|g_{uv}|^2P_u}{8\ln \frac{1.25}{\delta}+|N(v)|{\epsilon_v^*}^2}}, \text{if }  \epsilon_v^*\leq\epsilon_v^{(1)},\\
			\min_{u\in N(v)}\sqrt{|g_{uv}|^2P_u}, \text{if }  \epsilon_v^{(1)} \leq \epsilon_v^*\leq \epsilon_v^{(0)},
		\end{array}
		\right.
	\end{aligned}
\label{c2}
\end{equation}
\end{small}where \begin{footnotesize}
$$\epsilon_v^{(1)}\triangleq\sqrt{\frac{{8\ln \frac{1.25}{\delta}\min_{u\in N(v)}|g_{uv}|^2P_u}}{\sum_{u\in N(v)}|g_{uv}|^2P_u+\sigma_v^2-|N(v)|\min_{u\in N(v)}|g_{uv}|^2P_u}}.$$\end{footnotesize}To be more specific, when $\epsilon_v^*\leq\epsilon_v^{(1)}$ holds, by putting the value of $C_v$ into (\ref{rset3}), the values of  $\{\beta_{uv}\}_{u\in N(v)}$ are given by
$\beta_{uv}=1-C_v^2/(|g_{uv}|^2P_u), \forall u\in N(v).$
On the other hand, when $\epsilon_v^{(1)} \leq \epsilon_v^*\leq \epsilon_v^{(0)}$ holds, the values of  $\{\beta_{uv}\}_{u\in N(v)}$ should satisfy\begin{footnotesize}
\begin{subnumcases}{}
	0 \leq \beta_{uv}\leq 1-\frac{\min_{u\in N(v)}|g_{uv}|^2P_u}{|g_{uv}|^2P_u} \triangleq \beta_{uv}^U, \forall u\in N(v), &\label{bset1} \\
	\sum_{u\in N(v)}|g_{uv}|^2\beta_{uv}P_u=\frac{8\ln \frac{1.25}{\delta}\min_{u\in N(v)}|g_{uv}|^2P_u}{{\epsilon_v^*}^2}-\sigma_v^2,&\label{bset2}
\end{subnumcases}
\end{footnotesize}and can be solved by a water-filling alike algorithm summarized as Algorithm I. The key idea is allocating the right-hand-side of  (\ref{bset2}) among $|N(v)|$ neighbors, where $\mathbbm{1}_A$ is the indicator function defined as follows
\begin{align}
	\mathbbm{1}_A=\left\{\begin{array}{rcl}
		1, &\text{if event } A \text{ occurs},  \\
		0,  &\text{otherwise}.\\
	\end{array} \right. \label{feasible_phi}
\end{align}

\begin{algorithm}[]
	\caption{Water-Filling Alike Algorithm for $\{\beta_{uv}\}_{u\in N(v)}$}
	\label{A1}
	{\small
		\begin{algorithmic}[1]
			\STATE \textbf{initialization}
			\begin{itemize}
				\item Set $D = \frac{8\ln \frac{1.25}{\delta}\min_{u\in N(v)}|g_{uv}|^2P_u}{{\epsilon_v^*}^2}-\sigma_v^2$, $I = N(v)$.
			\end{itemize}
			\WHILE{$D > 0$}
			\STATE \textbf{Computing Water-filling Threshold:} $\bar{D}=\frac{D}{|I|}$.
			\STATE \textbf{Deciding Filling Strategy:}
			\IF {$|g_{uv}|^2\beta_{uv}^UP_u \geq \bar{D}, \forall u\in I$} 
			\STATE Set $\beta_{uv}=\frac{\bar{D}}{|g_{uv}|^2P_u}, \forall u\in I.$
			\STATE Set $D=0$.
			\ELSE
			\STATE Set $\beta_{uv}=\beta_{uv}^U\mathbbm{1}_{|g_{uv}|^2\beta_{uv}^UP_u \leq \bar{D}}, \forall u\in I$.
			\STATE Update $D = D- \sum_{u\in I} |g_{uv}|^2\beta_{uv}^UP_u\mathbbm{1}_{|g_{uv}|^2\beta_{uv}^UP_u \leq \bar{D}}.$
			\STATE Pop out $u$ with $ |g_{uv}|^2\beta_{uv}^UP_u \leq \bar{D}$ from $I$.
			\ENDIF
			\ENDWHILE
	\end{algorithmic}}
	\label{alg1}
\end{algorithm}

Based on the above analysis, the optimal solutions for Problem (\ref{Design_problem}) can be summarized as follows 
\begin{equation}
	\begin{aligned}
		\alpha_{uv} 
		&= \left\{
		\begin{array}{lr}
			\frac{{\epsilon_v^*}^2(\sigma_v^2+\sum_{u\in N(v)}|g_{uv}|^2P_u)}{|g_{uv}|^2P_u(8\ln \frac{1.25}{\delta}+|N(v)|{\epsilon_v^*}^2)}, \text{if }  \epsilon_v^*\leq\epsilon_v^{(1)},\\
			\frac{\min_{u\in N(v)}|g_{uv}|^2P_u}{|g_{uv}|^2P_u}, \text{if }  \epsilon_v^*> \epsilon_v^{(1)}.
		\end{array}
		\right.
	\end{aligned}
	\label{alpha_final}
\end{equation}
\begin{equation}
	\begin{aligned}
		\beta_{uv} 
		&= \left\{
		\begin{array}{lr}
			1-\alpha_{uv}, \text{if }  \epsilon_v^*\leq\epsilon_v^{(1)},\\
			\text{given by Algorithm I}, \text{ if }  \epsilon_v^{(1)} \leq \epsilon_v^*\leq \epsilon_v^{(0)},\\
			0, \text{if } \epsilon_v^* > \epsilon_v^{(0)}.
		\end{array}
		\right.
	\end{aligned}
	\label{beta_final}
\end{equation}

\subsection{SNR-Privacy Trade-off and Performance Analysis} \label{section:tradeoff}
In this part, we analyze the performance of the proposed privacy-preserving wireless signaling. Specifically, we focus on the trade-off between the SNR of $\bm{\tilde{R}}_v^{(1)}$ and the achieved privacy guarantee at the first iteration when predicting node $v$'s result. These two metrics influence the inference performance and privacy, respectively. With the solutions in (\ref{alpha_final}) and (\ref{beta_final}), the SNR of the received signal $\bm{\tilde{R}}_v^{(1)}$ is maximized while $\epsilon_v^*$ is satisfied and the maximum value  is given by
\begin{equation}
	\begin{aligned}
		\rho_v^{\max} 
		&= \left\{
		\begin{array}{lr}
			\frac{{\epsilon_v^*}^2}{8\ln \frac{1.25}{\delta}}, \text{if }  \epsilon_v^*\leq\epsilon_v^{(0)},\\
			\frac{\min_{u\in N(v)}|g_{uv}|^2P_u}{\sigma_v^2}, \text{if } \epsilon_v^* > \epsilon_v^{(0)},
		\end{array}
		\right.
	\end{aligned}
	\label{rho_final}
\end{equation}
which serves as the upper bound of the received SNR with a given privacy requirement, i.e., a given target privacy budget. In the following, we define the SNR-privacy trade-off function and the achievable (SNR, privacy guarantee) pair.

{\bf{Definition 3:}} (SNR-Privacy Trade-off Function) During decentralized inference with GNNs in wireless networks, the SNR-privacy trade-off function of node $v$ is given by
\begin{equation}
	\begin{aligned}
		\rho_v^{\max}(\epsilon_v)
		&= \left\{
		\begin{array}{lr}
			\frac{\epsilon_v^2}{8\ln \frac{1.25}{\delta}}, \text{if }  \epsilon_v\leq\epsilon_v^{(0)},\\
			\frac{\min_{u\in N(v)}|g_{uv}|^2P_u}{\sigma_v^2}, \text{if } \epsilon_v > \epsilon_v^{(0)},
		\end{array}
		\right.
	\end{aligned}
	\label{snr-privacy-function}
\end{equation}
where a ($\rho_v, \epsilon_v$) pair is defined as achievable if $\rho_v \leq \rho_v^{\max} (\epsilon_v)$ holds. Note that the statement that a ($\rho_v, \epsilon_v$) pair is achievable means that there exists at least a signal design solution, i.e., $\{\alpha_{uv},\beta_{uv}\}_{u\in N(v)}$, such that the SNR of  $\bm{\tilde{R}}_v^{(1)}$ is $\rho_v$ while $(\epsilon_v, \delta)$-LDP is achieved.

\begin{figure}
	\vspace{-2em}
	\centering	\includegraphics[width=0.9\linewidth, height=0.2\textheight]{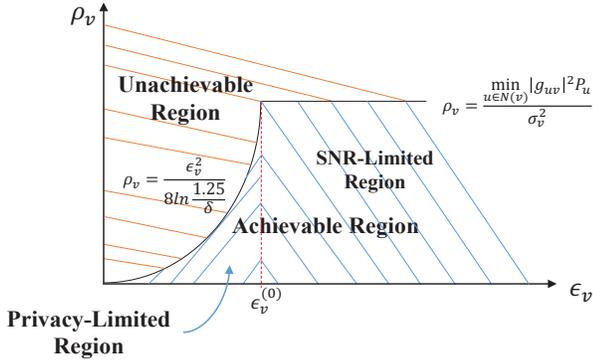}
	\vspace{-1em}
	\caption{A graphical illustration of the SNR-privacy trade-off function.}
	\label{fig:tradeoff}
	\vspace{-1em}
\end{figure}

The graphical illustration of the SNR-privacy trade-off function is shown in Fig. \ref{fig:tradeoff}. As suggested by the graph, the SNR-privacy trade-off function divides the whole space into achievable and unachievable regions, which are marked by blue and orange slashes, respectively. Moreover, the achievable region can be further divided into two sub-regions. Specifically, when $\epsilon_v\leq\epsilon_v^{(0)}$ holds, the upper bound of $\rho_v$ only depends on the hyperparameters of privacy requirements, i.e., $\epsilon_v$ and $\delta$. From Fig. \ref{fig:tradeoff}, the upper bound of the received SNR increases with $\epsilon_v$, which indicates that the performance of decentralized inference with GNNs can be enhanced by the loss of privacy. Therefore, this sub-region is termed as privacy-limited region, which is marked with blue forward slashes. On the other hand, when $\epsilon_v>\epsilon_v^{(0)}$ holds, the upper bound of $\rho_v$ only depends on the hyperparameters of wireless communications, including $\sigma_v^2$ and the worst-case received power from neighbors, i.e, $\min_{u\in N(v)}|g_{uv}|^2P_u$. Increasing $\epsilon_v$ in this sub-region does not lead to SNR gain, and the performance of decentralized inference with GNNs can only be enhanced by more abundant wireless resources. Therefore, this sub-region is termed as SNR-limited region, which is marked with blue backslashes. The above analysis will be further validated by the simulation results in Section \ref{testing:tradeoff}.

\section{ Privacy-Guaranteed Training Algorithm}
\label{section:training}
To achieve privacy-preserving decentralized inference, both artificial and channel noise are exploited in the wireless signaling proposed above. Unfortunately, they may deteriorate the estimation accuracy on $\{\bm{\hat{h}}_v^{(k)}\}_{k=1}^K$, and therefore cause inference performance loss, such as lower accuracy or higher mean squared error,  compared to the case with perfect wireless transmission and no privacy guarantee. Although we maximize the SNR of $\{\bm{\tilde{R}}_v^{(k)}\}_{k=1}^K$ with the solutions in (\ref{final_noprivacy}), (\ref{alpha_final}), and (\ref{beta_final}) to enhance the estimation accuracy on $\{\bm{\hat{h}}_v^{(k)}\}_{k=1}^K$, the inference performance loss induced by noise and fading still exists. Traditional training algorithms, however, do not take wireless environments into consideration and cannot overcome the above issue. In this section, we modify the training process of GNNs based on the privacy-preserving wireless signaling mentioned in Section \ref{section:signal} to mitigate the impact of noise and fading and enhance the inference performance of decentralized GNNs under a certain privacy guarantee.

The basic idea is training with noise and fading, where noise includes both artificial and channel noise.  Compared with the traditional training algorithm of GNNs, there are three improvements/differences in the proposed privacy-guaranteed training algorithm as follows.

{\bf1. Enhanced training samples:} For the traditional training algorithm, a training sample corresponds to a graph data, including its graph adjacency matrix and the node/edge features.	However, in our proposed training algorithm, parameters of wireless transmissions and privacy preservation are also included in a training sample. Specifically, parameters of wireless transmissions consist of channel coefficients $g_{uv}$, channel noise variances $\sigma_v^2$, and transmit power limits $P_v$ for all users, while parameters of privacy preservation refer to $\epsilon_v^*$ and $\delta$.

{\bf2. Pre-processing on training samples:} Before training, $\{\gamma_{uv},\alpha_{uv},\beta_{uv}\}$  is computed based on (\ref{final_noprivacy}), (\ref{alpha_final}), and (\ref{beta_final}) for each training sample.

{\bf3. Modified forward pass:} In the traditional training algorithm, the forward pass refers to the computation and storage of intermediate variables and outputs following the operations defined in (\ref{sum_tran}) and (\ref{sum_out}) from the first layer to the $K$-th layer. To resist the influence of noise and fading, we inject noise and fading during the forward pass in our proposed training algorithm by following the idea of adversarial training. Therefore, the operations in (\ref{sum_tran}) during the forward pass are modified as (\ref{comp1}).
\begin{figure*}[tb]
	\small
\begin{equation}
	\begin{aligned}
		\bm{\hat{h}}_v^{(k)} 
		&= \left\{
		\begin{array}{lr}
			\sum_{u\in N(v)}[\check{f}_M^{(1)}(\bm{h}_u^{(0)},\bm{e}_{vu})+\frac{|g_{uv}|}{C_v}\sqrt{\beta_{uv}P_u}\bm{m}_{uv}]+\frac{\bm{n}_v^{(1)}}{C_v}, \text{if }  k=1,\\
			\sum_{u\in N(v)}f_M^{(k)}(\bm{h}_u^{(k-1)},\bm{e}_{vu})+\frac{\bm{n}_v^{(k)}}{\sqrt{\min_{u\in N(v)}|g_{uv}|^2P_u}}, \text{otherwise}.
		\end{array}
		\right.
	\end{aligned}
	\label{comp1}
\end{equation}
	\hrulefill
\end{figure*}
The backward gradient propagation keeps the same as the traditional training algorithm and therefore the convergence of the proposed training algorithm can be guaranteed. 

\begin{figure}
	\centering
\includegraphics[width=0.98\linewidth, height=0.25\textheight]{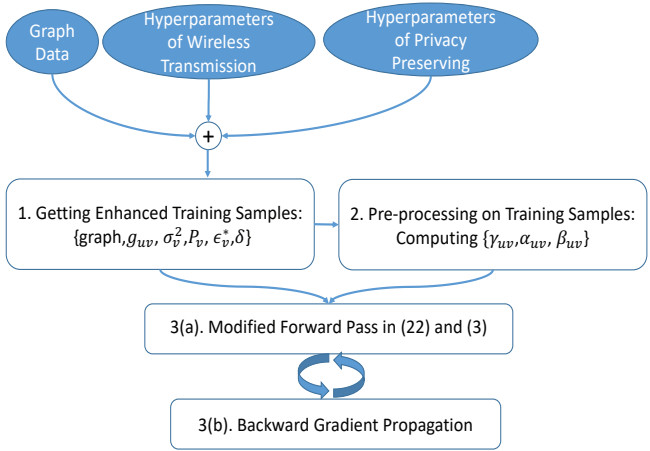}
	\vspace{-1em}
	\caption{A graphical illustration of the  privacy-guaranteed training algorithm.}
	\label{fig:training}
\end{figure}

The main process of the proposed privacy-guaranteed training algorithm is summarized in Fig. \ref{fig:training}, whose advantage will be validated by the simulation results in Section \ref{test:training}.

\section{Extension to systems without Aircomp} \label{section:aircomp}
This paper thus far proposes the optimal wireless signal design, performance upper bound analysis, and modified training algorithms for privacy-preserving decentralized inference with GNNs. All above mentioned design and analysis have universality, which are not only applicable for different wireless applications but also for different wireless communication models. On the other hand, the Aircomp technique sometimes may fail since it relies on the high-accurate synchronization among users\footnote{Node synchronization technique is out of the scope of this paper but has been well studied in literature, such as \cite{sync1,sync2}.}. Moreover, some aggregation functions of GNNs, such as  $\max(\cdot)$, cannot be achieved by the Aircomp technique. To not only validate the universality of our design and analysis but also avoid the possible system failure induced by the Aircomp technique, in this section, we extend our design and analysis to the wireless system without the Aircomp technique,  and meanwhile further emphasize the superiority of the Aircomp technique for privacy-preserving decentralized inference with GNNs.

\subsection{Wireless Communication Model}
Without the Aircomp technique, we consider a single-antenna orthogonal system with slow fading and i.i.d Gaussian noise. The CSI is still available to both the transmitter and receiver. During decentralized inference with GNNs, the received signal of node $v$ from its neighbor $u\in N(v)$ at the $k$-th iteration, denoted as $\bm{R}_{uv}^{(k)}$, is given by
\begin{equation*}
	\begin{aligned}
		\bm{R}_{uv}^{(k)} &= g_{uv}\bm{w}_{uv}^{(k)}+\bm{n}_{uv}^{(k)}=  g_{uv}\sqrt{P_u}f_M^{(k)}(\bm{h}_u^{(k-1)},\bm{e}_{vu})+\bm{n}_{uv}^{(k)},
	\end{aligned}
\end{equation*}
where the elements in $\bm{n}_{uv}^{(k)}$ are i.i.d drawn from $\mathcal{N}(0,\sigma_v^2)$. Similar to the analysis in Section \ref{section:communication_model}, directly transmitting $\bm{w}_{uv}^{(k)}$ makes it difficult to achieve privacy-preserving inference, which indicates the necessity of novel wireless signaling.

\subsection{Privacy-Preserving Wireless Signaling}\label{section:comp_analysis}
Similar to the analysis in Section \ref{section:signal}, the transmitting signals for the last $K-1$ iterations only aim at accurate estimations on $\{f_M^{(k)}(\bm{h}_u^{(k-1)},\bm{e}_{vu})\}_{k=2}^K$. To achieve this goal, we can adopt the signal proposed in (\ref{each_user_1}) and setting $\gamma_{uv}=1$ since there is no amplitude alignment constraint for the system without the Aircomp technique.

On the other hand, for the first iteration, the transmitting signal should provide privacy guarantee. Therefore, we can adopt the signal designed in (\ref{each_user_signal}), and the received signal of node $v$ from its neighbor $u\in N(v)$ at the first iteration, denoted as $\bm{\tilde{R}}_{uv}^{(1)}$, is given by
\begin{equation}
	\begin{aligned}
		\bm{\tilde{R}}_{uv}^{(1)}= & |g_{uv}|\sqrt{\alpha_{uv}P_u}\check{f}_M^{(1)}(\bm{h}_u^{(0)},\bm{e}_{vu})\\&+|g_{uv}|\sqrt{\beta_{uv}P_u}\bm{m}_{uv}+\bm{n}_{uv}^{(1)}.
	\end{aligned}
	\label{rv_no}
\end{equation}
We introduce the following theorem to analyze the privacy guarantee achieved by $\bm{\tilde{R}}_{uv}^{(1)}$ for node $u$, whose proof can be found in Appendix \ref{epsion2}.

{\bf{Theorem 2:}} For node $u$ in $N(v)$,   $\bm{\tilde{R}}_{uv}^{(1)}$ achieves $(\epsilon_{uv}, \delta)$-LDP when predicting node $v$'s result, where
$$\epsilon_{uv}=\frac{2|g_{uv}|\sqrt{\alpha_{uv}P_u}}{\sqrt{|g_{uv}|^2\beta_{uv}P_u+\sigma_v^2}}\sqrt{2\ln \frac{1.25}{\delta}}. $$

Furthermore, without the help of the Aircomp technique, node $v$ needs to first estimate $\check{f}_M^{(1)}(\bm{h}_u^{(0)},\bm{e}_{vu})$ from $\bm{\tilde{R}}_{uv}^{(1)}$ for all neighbors in $N(v)$, and then compute the sum of $\{\check{f}_M^{(1)}(\bm{h}_u^{(0)},\bm{e}_{vu})\}_{u \in N(v)}$ to estimate $\bm{\hat{h}}_v^{(1)}$ as follows
\begin{small}
\begin{equation}
	\begin{aligned}
	 \sum_{u\in N(v)}\frac{\bm{\tilde{R}}_{uv}^{(1)}}{|g_{uv}|\sqrt{\alpha_{uv}P_u}} 
		= &\sum_{u\in N(v)}\check{f}_M^{(1)}(\bm{h}_u^{(0)},\bm{e}_{vu}) \\&+ \sum_{u\in N(v)}[\sqrt{\frac{\beta_{uv}}{\alpha_{uv}}}\bm{m}_{uv}+\frac{\bm{n}_{uv}^{(1)}}{|g_{uv}|\sqrt{\alpha_{uv}P_u}}].\\
	\end{aligned}
\label{infer_no}
\end{equation}
\end{small}To enhance the accuracy of the estimation on $\bm{\hat{h}}_v^{(1)}$, the SNR of the estimation result in (\ref{infer_no}), defined as the ratio between the power of efficient signal and the noise, needs to be maximized. Specifically, the SNR of the estimation result in (\ref{infer_no}), denoted by $\hat{\rho}_v$, is mathematically given by
\begin{equation}
	\hat{\rho}_v=\frac{1}{\sum_{u\in N(v)}\frac{|g_{uv}|^2\beta_{uv}P_u+\sigma_v^2}{|g_{uv}|^2\alpha_{uv}P_u}}.
	\label{SNR_no}
\end{equation}

With above analysis, the signal design problem for the system without the Aircomp technique is formulated as
\begin{equation}\label{Design_problem_no}
	\hat{\rho}_v^{\max} \triangleq \max_{\{\alpha_{uv},\beta_{uv}\}_{u\in N(v)}}\hat{\rho}_v=\frac{1}{\sum_{u\in N(v)}\frac{|g_{uv}|^2\beta_{uv}P_u+\sigma_v^2}{|g_{uv}|^2\alpha_{uv}P_u}\triangleq \hat{\rho}_{uv}}, 
\end{equation}
subject to (\ref{cons1}) and
\begin{align}
	\frac{2|g_{uv}|\sqrt{\alpha_{uv}P_u}}{\sqrt{|g_{uv}|^2\beta_{uv}P_u+\sigma_v^2}}\sqrt{2\ln \frac{1.25}{\delta}} \leq \epsilon_v^*,  \forall u\in N(v), \label{cons3_no}\tag{\theequation a}
\end{align}
where $\hat{\rho}_{uv}$ is the SNR of $\bm{\tilde{R}}_{uv}^{(1)}$ defined as the ratio between the power of $\check{f}_M^{(1)}(\bm{h}_u^{(0)},\bm{e}_{vu})$ and the corresponding noise.

Note that Problem (\ref{Design_problem_no}) can be decomposed into $|N(v)|$ sub-problems since no amplitude alignment is needed and the optimal signal design is independent among different neighbors.  The solving process of each sub-problem is almost the same as that of Problem (\ref{Design_problem}) in Section \ref{section:optimal_solution}. Due to the page limit, we omit the detailed derivation but directly give out the optimal solutions of Problem (\ref{Design_problem_no}) as follows
\begin{equation}
	\begin{aligned}
		\alpha_{uv} 
		&= \left\{
		\begin{array}{lr}
			\frac{{\epsilon_v^*}^2(\sigma_v^2+|g_{uv}|^2P_u)}{|g_{uv}|^2P_u(8\ln \frac{1.25}{\delta}+{\epsilon_v^*}^2)}, \text{if }  \epsilon_v^*\leq\epsilon_{uv}^{(0)},\\
			1, \text{if }  \epsilon_v^*> \epsilon_{uv}^{(0)},
		\end{array}
		\right.
	\end{aligned}
	\label{alpha_final_no}
\end{equation}
\begin{equation}
	\begin{aligned}
		\beta_{uv} 
		&= 1-\alpha_{uv},
	\end{aligned}
	\label{beta_final_no}
\end{equation}
where $\epsilon_{uv}^{(0)}=\sqrt{8\ln \frac{1.25}{\delta}|g_{uv}|^2P_u/\sigma_v^2}.$

With the optimal signal design given by (\ref{alpha_final_no}) and (\ref{beta_final_no}), the SNR of the received signal $\bm{\tilde{R}}_{uv}^{(1)}$ and the SNR of the estimation result in (\ref{infer_no}) are simultaneously maximized while $\epsilon_v^*$ is satisfied, whose maximum values  are given by
\begin{equation*}
	\begin{aligned}
	\hat{\rho}_{uv}^{\max}
		&= \left\{
		\begin{array}{lr}
			\frac{{\epsilon_v^*}^2}{8\ln \frac{1.25}{\delta}}, \text{if }  \epsilon_v^*\leq\epsilon_{uv}^{(0)},\\
			\frac{|g_{uv}|^2P_u}{\sigma_v^2}, \text{if } \epsilon_v^* > \epsilon_{uv}^{(0)},
		\end{array}
	\text{and }
		\right.
		\hat{\rho}_v^{\max} = \frac{1}{\sum_{u\in N(v)\frac{1}{\hat{\rho}_{uv}^{\max}}}},
	\end{aligned}
	\label{rho_final_no}
\end{equation*}
respectively. Therefore, the SNR-privacy trade-off function of node $v$ is given by
\begin{equation}
	\begin{aligned}
		\hat{\rho}_v^{\max}(\epsilon_v) = \frac{1}{\sum_{u\in N(v)\frac{1}{\hat{\rho}_{uv}^{\max}(\epsilon_v)}}},
	\end{aligned}
\end{equation}
where
\begin{equation*}
	\begin{aligned}
		\hat{\rho}_{uv}^{\max}(\epsilon_v)
		&= \left\{
		\begin{array}{lr}
			\frac{\epsilon_v^2}{8\ln \frac{1.25}{\delta}}, \text{if }  \epsilon_v\leq\epsilon_{uv}^{(0)},\\
			\frac{|g_{uv}|^2P_u}{\sigma_v^2}, \text{if } \epsilon_v > \epsilon_{uv}^{(0)}.
		\end{array}
		\right.
	\end{aligned}
	\label{snr-privacy-function_no}
\end{equation*}

Compared with the SNR-privacy trade-off function defined in (\ref{snr-privacy-function}) for the system with the Aricomp technique, we find that $\rho_v^{\max}>\hat{\rho}_v^{\max}$ always holds. This conclusion indicates that adopting the Aricomp technique can enhance the SNR of the estimation result on $\bm{\hat{h}}_v^{(1)}$ during decentralized inference with GNNs while achieving the same privacy target. The reasons are as follows. Without
the Aircomp technique, all signals are transmitted in orthogonal and the privacy of each neighbor
is only preserved by the artificial and channel noise. However, in the system with the
Aircomp technique, the signals from all neighbors are mixed. Therefore, the privacy of each
neighbor is not only preserved by the artificial and channel noise, but also the transmitting
signals of other users. In this way, it is more difficult for a honest-but-curious neighbor or an adversary to successfully detect
the personal features of each user. Therefore, the Aircomp technique can not only enhance
communication and computation efficiency but also improve the performance/privacy of decentralized inference with
GNNs, which will be further validated by the simulation results in Section \ref{simulation:aircomp}.

\subsection{Privacy-Guaranteed Training Algorithm}
The privacy-guaranteed training algorithm for the system without the Aircomp technique is similar to that in Section \ref{section:training}. The only difference is that the operations in (\ref{comp1}) are modified as (\ref{tran1}).
\begin{figure*}[tb]
\small
\begin{equation}
\begin{aligned}
	\bm{\hat{h}}_v^{(k)} 
	&= \left\{
	\begin{array}{lr}
	\sum_{u\in N(v)}[\check{f}_M^{(1)}(\bm{h}_u^{(0)},\bm{e}_{vu})+\sqrt{\frac{\beta_{uv}}{\alpha_{uv}}}\bm{m}_{uv}+\frac{\bm{n}_{uv}^{(1)}}{|g_{uv}|\sqrt{\alpha_{uv}P_u}}], \text{if }  k=1,\\
	\sum_{u\in N(v)}[f_M^{(k)}(\bm{h}_u^{(k-1)},\bm{e}_{vu})+\frac{\bm{n}_{uv}^{(k)}}{\sqrt{|g_{uv}|^2P_u}}], \text{otherwise}.
	\end{array}
	\right.
\end{aligned}
\label{tran1}
\end{equation}
\hrulefill
\end{figure*}

\section{Simulation Results} \label{section:simulation}
In this section, we use simulation results to validate the efficiency of our proposed privacy-preserving wireless signaling and privacy-guaranteed training algorithm. Although our proposed mechanism is a general framework and can be applied for various wireless optimization problems, we use the sum rate maximization problem in the Gaussian interference channel as an example for the following simulation, which serves as a classic application for GNN based wireless techniques {\cite{graphnn,shen_graph}}. We first briefly introduce
the system model and graphical model of the sum rate maximization problem. Then we present the testing results for this application. 

\subsection{Sum Rate Maximization Problem} \label{testing:model}
\begin{figure}[h]
	\vspace{-2em}
	\centering
	\subfigure[Wireless system model.]{
		\begin{minipage}[t]{0.45\linewidth}
			\centering
			\includegraphics[width=1.5in]{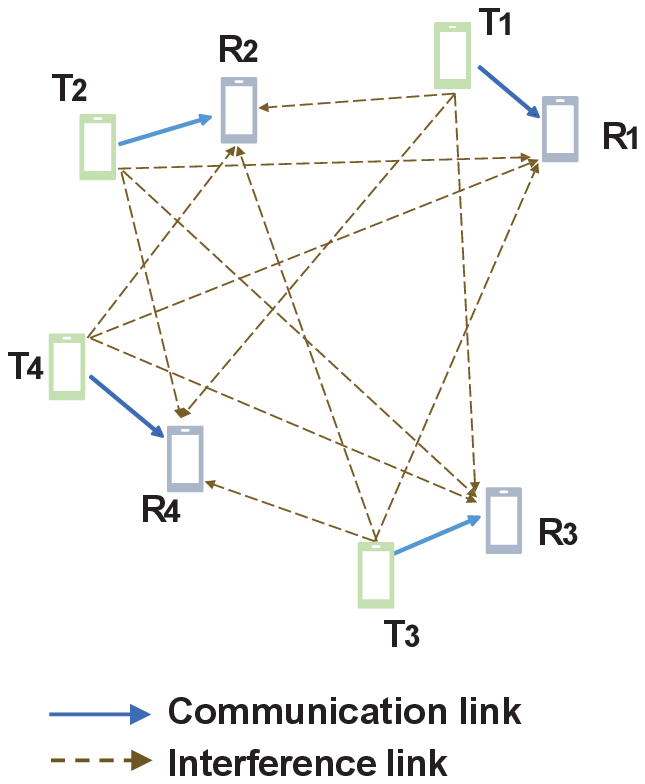}
			\label{fig:D2D}
		\end{minipage}%
	}
	\subfigure[Graphical model.]{
		\begin{minipage}[t]{0.45\linewidth}
			\centering
			\includegraphics[width=1.4in]{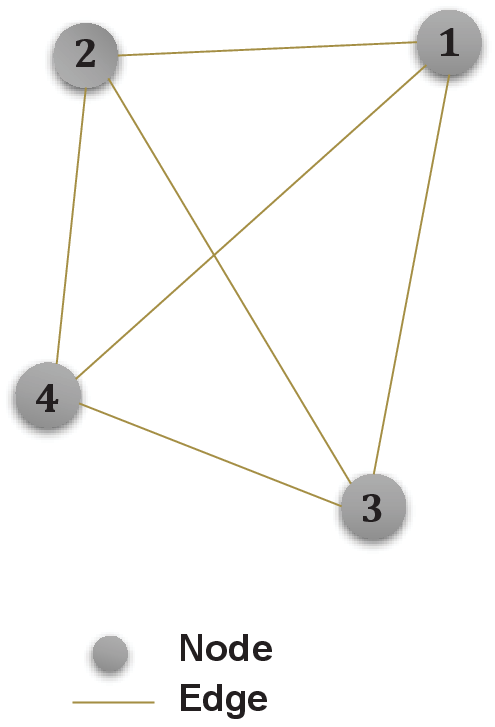}
			\label{fig:fully}
		\end{minipage}
	}
\vspace{-0.5em}
\caption{Wireless system model and graphical model for the sum rate maximization problem.}
\end{figure}

As shown in Fig. \ref{fig:D2D}, we consider a system with $N$ mutually interfering single-antenna device-to-device (D2D) pairs. Let $g_{j,i}$ denote the complex valued channel coefficient from transmitter $j$ to receiver $i$, which is drawn from the complex Gaussian distribution $\mathcal{CN}(0,1)$. With additive Gaussian noise i.i.d drawn from $\mathcal{N}(0,\sigma_i^2)$, the signal-to-interference-plus-noise ratio (SINR) for receiver $i$ is given by 
$SINR_i = |g_{i,i}|^2p_i/(\sum_{j\neq i}|g_{j,i}|^2p_j+\sigma_i^2)$,
where $p_i$ is the transmit power of transmitter $i$ and needs to be optimized to maximize the sum rate  of all $N$ receivers. In this way, the mathematical formulation of the sum rate maximization problem is given by
\begin{equation}\label{sumrate_problem}
	\max_{\{p_i\}_{i=1}^N} \sum_{i=1}^{N}\log(1+SINR_i), 
\end{equation}
subject to 
\vspace{-1em}
\begin{align}
	0\leq p_i \leq P^{\max}, i=1, 2, ... N, \label{rate}\tag{\theequation a}
\end{align}
where $P^{\max}$ is the maximum transmit power of each transmitter  for the sum rate maximization problem. 

Note that hyperparameters $\{g_{uv},\sigma_v^2,P_v\}$ for decentralized inference with GNNs have different physical meanings from hyperparameters $\{g_{i,j},\sigma_i^2,P^{\max}\}$ in the sum rate maximization problem (\ref{sumrate_problem}). The former ones are for decentralized control/management with GNNs while the latter ones are for the subsequent transmission after decentralized control/management. However, in the simulation throughout this paper, $g_{i,j}=g_{vu}$ and $\sigma_v^2=\sigma_i^2$ hold since the wireless environment remains the same for the control and transmission stages.

The graphical model of the sum rate maximization problem in (\ref{sumrate_problem}) is shown as Fig. \ref{fig:fully}, where each node corresponds to a D2D pair and two nodes are connected by an edge if the corresponding D2D pairs interfere with each other. Since all D2D pairs are mutually interfering with each other, the corresponding graphical model is fully-connected. For each node $i$, the node features\footnote{If node features are detected, hackers can transmit noise with certain power (calculated with detected parameters) to interfere with the following communication and crash the communication system, which indicates the necessity of privacy preservation.} consist of $|g_{i,i}|$ and $\sigma_i^2$. And for each edge between node $i$ and node $j$, the edge features refer to $|g_{i,j}|$ and $|g_{j,i}|$. With this graphical model, the sum rate maximization problem is converted to a node-level regression problem over the graph in  Fig. \ref{fig:fully}.

\subsection{Simulation Setup}
All codes are implemented in python 3.9, which run on the Ubuntu system with two 16-core Intel 2.90GHz CPUs and one GeForce RTX 3080.
\subsubsection{Parameters for Sum Rate Maximization Problem}
We consider a system with $N=10$ D2D pairs, and set $P^{\max}=30$ dBm and $\sigma_i=1$ for all transmitters.

\subsubsection{Parameters for Decentralized Inference with GNNs}
We adopt a 3-layer GNN for the sum rate maximization problem. All $\{f_M^{(k)},f_U^{(k)}\}_{k=1}^3$ are realized through MLPs, whose structures\footnote{\{a,b,c\} in Table \ref{table:mlp} indicates that the corresponding MLP has 3 layers, and the number of neurons at each layer is a, b, and c, respectively.} are summarized in Table \ref{table:mlp}. Moreover, we adopt the ReLU function as the activation function for each layer in the MLPs except the output layer of $f_U^{(3)}(\cdot)$, where the Sigmoid function serves as the activation function. 
\begin{table}
	\vspace{-2em}
	\footnotesize
	\caption{The Structure of MLPs for Local Message and Update Functions in GNNs}
	\vspace{-1em}
	\label{table:mlp}
	\centering
	\begin{tabular}{|c|c|}
		\hline
		Function & Number of Neurons in Corresponding MLP \\
		\hline
		$f_M^{(1)}$ & \{4,16,32\} \\
		\hline
		$f_M^{(2)}$ & \{34,64,32\} \\
		\hline
		$f_M^{(3)}$ & \{34,64,32\} \\
		\hline
		$f_U^{(1)}$ & \{34,16,32\} \\
		\hline
		$f_U^{(2)}$ & \{64,64,32\} \\
		\hline
		$f_U^{(3)}$ & \{64,64,16,1\} \\
		\hline
	\end{tabular}
\vspace{-1em}
\end{table}

During the training process of the GNN model, we use the negative of the sum rate as the loss function. Therefore, the GNN model is trained in the unsupervised manner, where labeled training samples that are difficult to obtain in wireless networks are not needed. Throughout the simulation, 10,000 unlabeled network layouts are used for training with the batch size as 64. According to \cite{shen_graph}, each unlabeled network layout is a fully-connected graph with 10 nodes and the corresponding complex valued channel coefficients $\{g_{i,j}\}$ are generated by following  $\mathcal{CN}(0,1)$. Moreover, the Adam optimizer \cite{adam} is adopted with a learning rate of 0.001. We also use batch normalization to avoid vanishing gradient issue. The training process will come to an end after 400 training epochs. 

As for the inference stage of the GNN model, 1,000 network layouts are used and all tables/figures in this section report the average performance over all testing network layouts. Meanwhile, unless otherwise stated, the privacy requirement is set as ($\epsilon_v^*=1, \delta=10^{-4}$)-LDP and the maximum transmit power for inference is set as $P_v=10$ dBm for all users. Furthermore, as mentioned above, $g_{i,j}=g_{vu}$ while $\sigma_v^2=\sigma_i^2=1$ throughout the simulation in this section.

\subsection{Performance of Privacy-Guaranteed Training Algorithm} \label{test:training}
In this part, we focus on the performance of the proposed privacy-guaranteed training algorithm in Section \ref{section:training}. We consider the following benchmarks for comparison\footnote{No state-of-the-art techniques are available for our considered problem. The simulation is conducted by comparing with some basic benchmarks.}.
\begin{itemize}
	\item Classic training algorithm: The forward pass  follows the operations defined in (\ref{sum_tran}) and (\ref{sum_out}), where fading and noise, including artificial and channel noise, are not considered during the training process.
	\item Training without artificial noise: The forward pass only takes channel fading and channel noise into consideration, where operations in  (\ref{sum_tran}) are modified as 
	\begin{small}$$\bm{\hat{h}}_v^{(k)} = \sum_{u\in N(v)}f_M^{(k)}(\bm{h}_u^{(k-1)},\bm{e}_{vu})+\frac{\bm{n}_v^{(k)}}{\min_{u\in N(v)}\sqrt{|g_{uv}|^2P_u}}, \forall k.$$\end{small}We consider this training algorithm because it only depends on hyperparameters of wireless transmission \{$g_{uv},\sigma_v^2,P_v$\}, which can be estimated from $\{g_{i,j},\sigma_i^2,P^{\max}\}$, i.e., the node and edge features of the graphical model for the sum rate maximization problem as introduced in Section \ref{testing:model}. Therefore, this training algorithm is feasible in practice with minor signaling and computation overhead than the privacy-guaranteed training algorithm.
\end{itemize}

The testing results are summarized in Table \ref{table:training}, where sum rates are normalized by the sum rate of running the classic weighted minimum mean square error (WMMSE) algorithm \cite{wmmse} for 100 iterations. In literature, WMMSE algorithm generally serves as a good performance upper bound for the sum rate maximization problem with interfering users \cite{shi,shen_graph}. Therefore, we use the normalized sum rate as a performance metric in the remainder of this paper.
\begin{table}
	\vspace{-2em}
	\footnotesize
	\caption{Performance of Different Training Algorithms}
	\vspace{-1em}
	\label{table:training}
	\centering
	\begin{tabular}{|c|c|}
		\hline
		Training Algorithm & Normalized Sum Rate \\
		\hline
		Classic training & 0.3701 \\
		\hline
		Training without artificial noise & 0.9549\\
		\hline
		Privacy-guaranteed training & 0.9587\\
		\hline
	\end{tabular}
\vspace{-2em}
\end{table}

From Table \ref{table:training}, our proposed privacy-guaranteed training algorithm outperforms both benchmarks, especially the classic training algorithm. This indicates the effectiveness of our proposed training algorithm. Another interesting finding is that the performance achieved by the training algorithm without the artificial noise has minor loss (less than 1\%) compared with the proposed privacy-guaranteed training algorithm. This indicates that we can adopt the training algorithm without artificial noise when the signaling/computation resource is limited in practice, which leads to marginal performance loss.

\subsection{Validation on SNR-Privacy Trade-off} \label{testing:tradeoff}
In this part, we discuss and validate the SNR-privacy trade-off defined in Section \ref{section:tradeoff}. Specifically, we pay attention to three metrics in this part. The first metric is the normalized sum rate introduced above. The second one is the percentage of users in the privacy-limited region. The final one is the SNR of the received signal $\bm{\tilde{R}}_v^{(1)}$ in (\ref{rv}) during the first iteration of the inference stage, whose value is normalized by corresponding maximum\footnote{Normalizing the SNR is for better graphical illustration.}.

\begin{figure}[h]
	\vspace{-1em}
	\centering
	\subfigure[SNR-privacy trade-off.]{
		\begin{minipage}[t]{1\linewidth}
			\centering
			\includegraphics[width=2.4in]{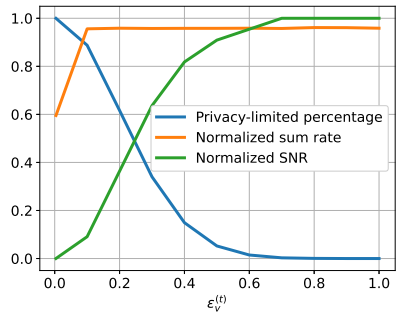}
			\label{fig:epsilon}
		\end{minipage}%
	}
	\subfigure[The impact of the maximum transmit power for inference.]{
		\begin{minipage}[t]{1\linewidth}
			\centering
			\includegraphics[width=2.4in]{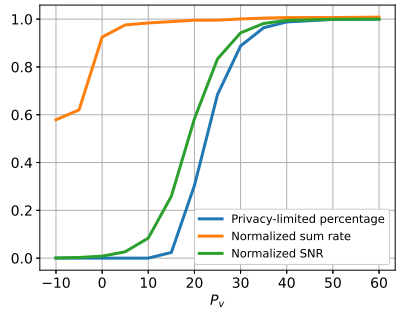}
			\label{fig:power}
		\end{minipage}
	}
	\vspace{-0.5em}
	\caption{SNR-privacy trade-off for decentralized inference with GNNs in wireless networks.}
	\vspace{-1em}
\end{figure}
To validate the SNR-privacy trade-off, we adopt the privacy-guaranteed training algorithm and change the value of the target privacy budget. The testing results are summarized in Fig. \ref{fig:epsilon}. From the graph, the tendency of the received SNR changing with the target privacy budget matches with our theoretical analysis  in Fig. \ref{fig:tradeoff}. Meanwhile, the normalized sum rate first increases and then fluctuates around a certain value with the increase of $\epsilon_v^*$. This result validates our discussion in Section \ref{section:tradeoff}. We draw the conclusion that increasing the target privacy budget to enhance the inference performance only holds when the  target privacy budget is smaller than a threshold, where most users are in the privacy-limited region. Otherwise, increasing target privacy budget only leads to weaker privacy guarantee but no performance gain, where most users are in the SNR-limited region. 

As analyzed in Section \ref{section:tradeoff}, the performance of decentralized inference with GNNs for users in the SNR-limited region can only be enhanced by more abundant wireless resources. This conclusion is then validated by the simulation results in Fig. \ref{fig:power}, where we conduct simulation with different maximum transmit powers for inference, i.e, $P_v$. From this figure, when the percentage of users in the privacy-limited region is small, most users are in the SNR-limited region, where increasing the maximum transmit power for inference can enhance the received SNR and the normalized sum rate simultaneously. The simulation results also indicate that increasing the maximum transmit power for inference over certain threshold (around 40 dBm in our simulation) makes all users in the privacy-limited region, which leads to no inference performance gain but only larger consumption on wireless resources. This conclusion sheds lights on power control for practical systems.

Furthermore, from both  Figs. \ref{fig:epsilon} and \ref{fig:power}, the normalized sum rate reaches the maximum value much faster than the received SNR. The reason is that GNN itself has some robustness/tolerance to the fading and noise as demonstrated by \cite{tmc}. The results also indicate that maximizing the received SNR during the inference stage is sufficient for optimizing the inference performance, which shows the rationality of our formulated problem (\ref{Design_problem}). 

\subsection{Performance of  Privacy-Preserving Wireless Signaling}
\begin{figure}[h]
	\centering
	\subfigure[Normalized SNR.]{
		\begin{minipage}[t]{1\linewidth}
			\centering
			\includegraphics[width=2.4in]{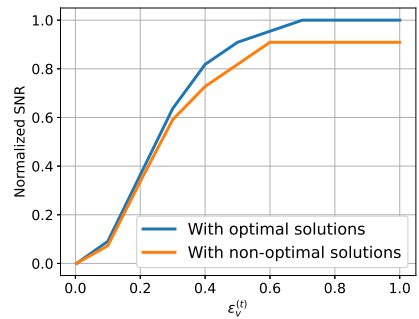}
			\label{fig:snr_opt}
		\end{minipage}%
	}
	\subfigure[Normalized sum rate.]{
		\begin{minipage}[t]{1\linewidth}
			\centering
			\includegraphics[width=2.4in]{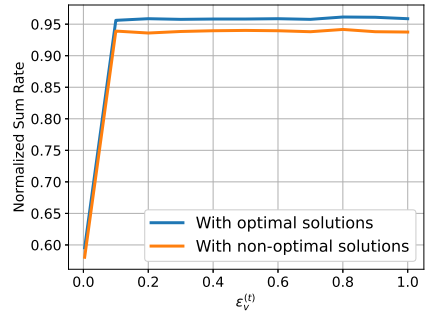}
			\label{fig:sumrate_opt}
		\end{minipage}
	}
	\vspace{-0.5em}
	\caption{Performance comparison between systems on privacy-preserving wireless signaling with optimal  and non-optimal solutions.}
	\label{fig:opt}
\end{figure}
In this part, we aim to verify the performance of our proposed privacy-preserving wireless signaling and conduct simulation on the following two signal design settings.
\begin{itemize}
	\item Privacy-preserving wireless signaling with optimal solutions given by (\ref{final_noprivacy}), (\ref{alpha_final}) and (\ref{beta_final}). 
	\item Privacy-preserving wireless signaling with non-optimal solutions by replacing (\ref{final_noprivacy}) and (\ref{beta_final}) with $\gamma_{uv}=\frac{\min_{u\in N(v)}|g_{uv}|^2P_u}{2|g_{uv}|^2P_u}$ and $\beta_{uv}=1-\alpha_{uv},\forall \epsilon_v^*$, respectively.
\end{itemize}

The simulation results are summarized in Fig. \ref{fig:opt}. When $\epsilon_v^*$ is small, $\beta_{uv}=1-\alpha_{uv}$ is the optimal solution according to (\ref{beta_final}). Therefore, there is no performance difference between the above two signal design settings in terms of the received SNR of $\bm{\tilde{R}}_v^{(1)}$ in (\ref{rv}). However, with the increase of $\epsilon_v^*$, the differences between the optimal and non-optimal solutions gradually become large, which causes performance loss on both the received SNR and the normalized sum rate when using the privacy-preserving signal with non-optimal solutions. The simulation results indicate the necessity and  superiority of the proposed privacy-preserving wireless signaling especially in the SNR-limited region.

\subsection{Validation on the Advantage of Aircomp} \label{simulation:aircomp}
\begin{figure}[h]
	\centering
	\subfigure[Normalized SNR.]{
		\begin{minipage}[t]{1\linewidth}
			\centering
			\includegraphics[width=2.4in]{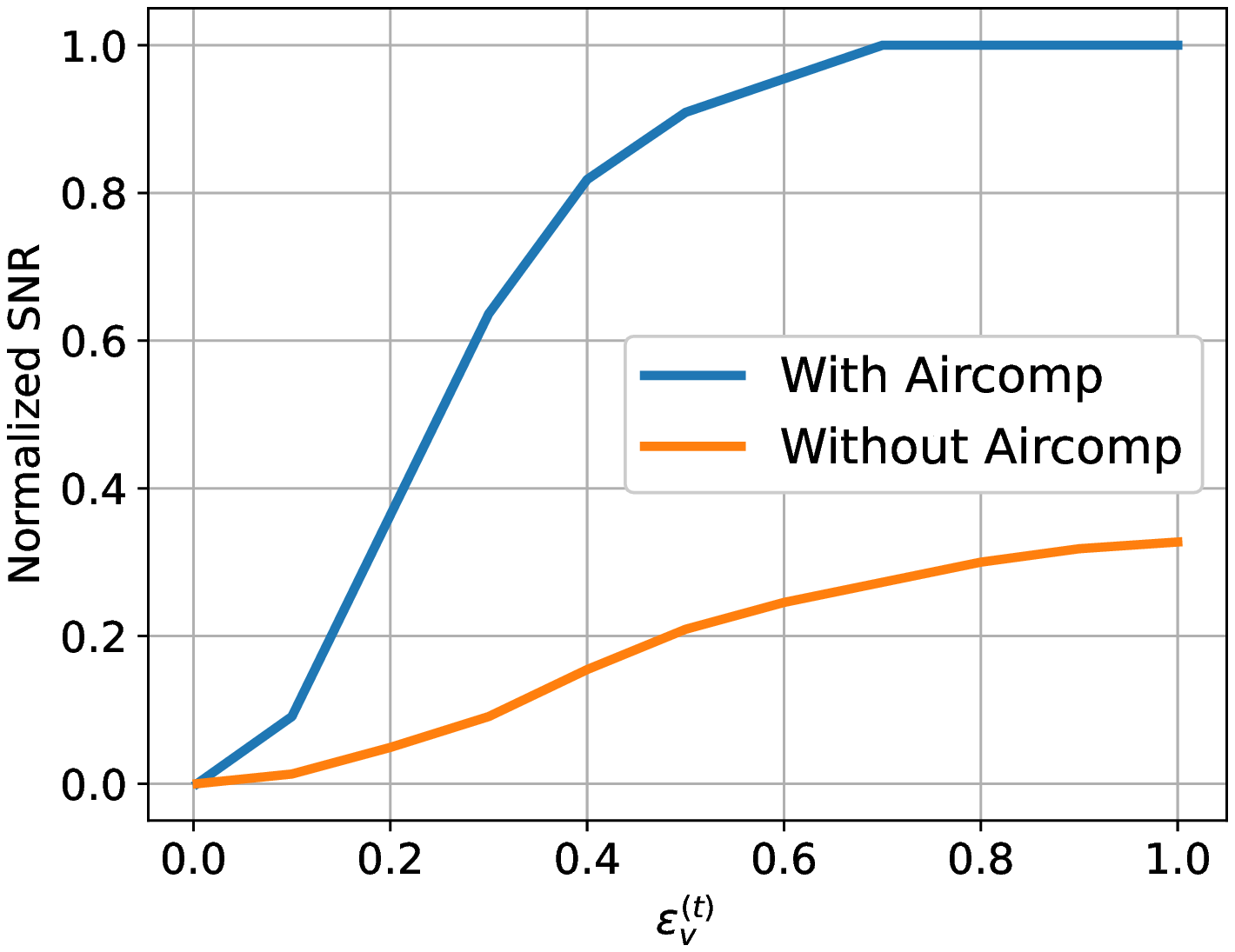}
			\label{fig:snr_comp}
		\end{minipage}%
	}
	\subfigure[Normalized sum rate.]{
		\begin{minipage}[t]{1\linewidth}
			\centering
			\includegraphics[width=2.4in]{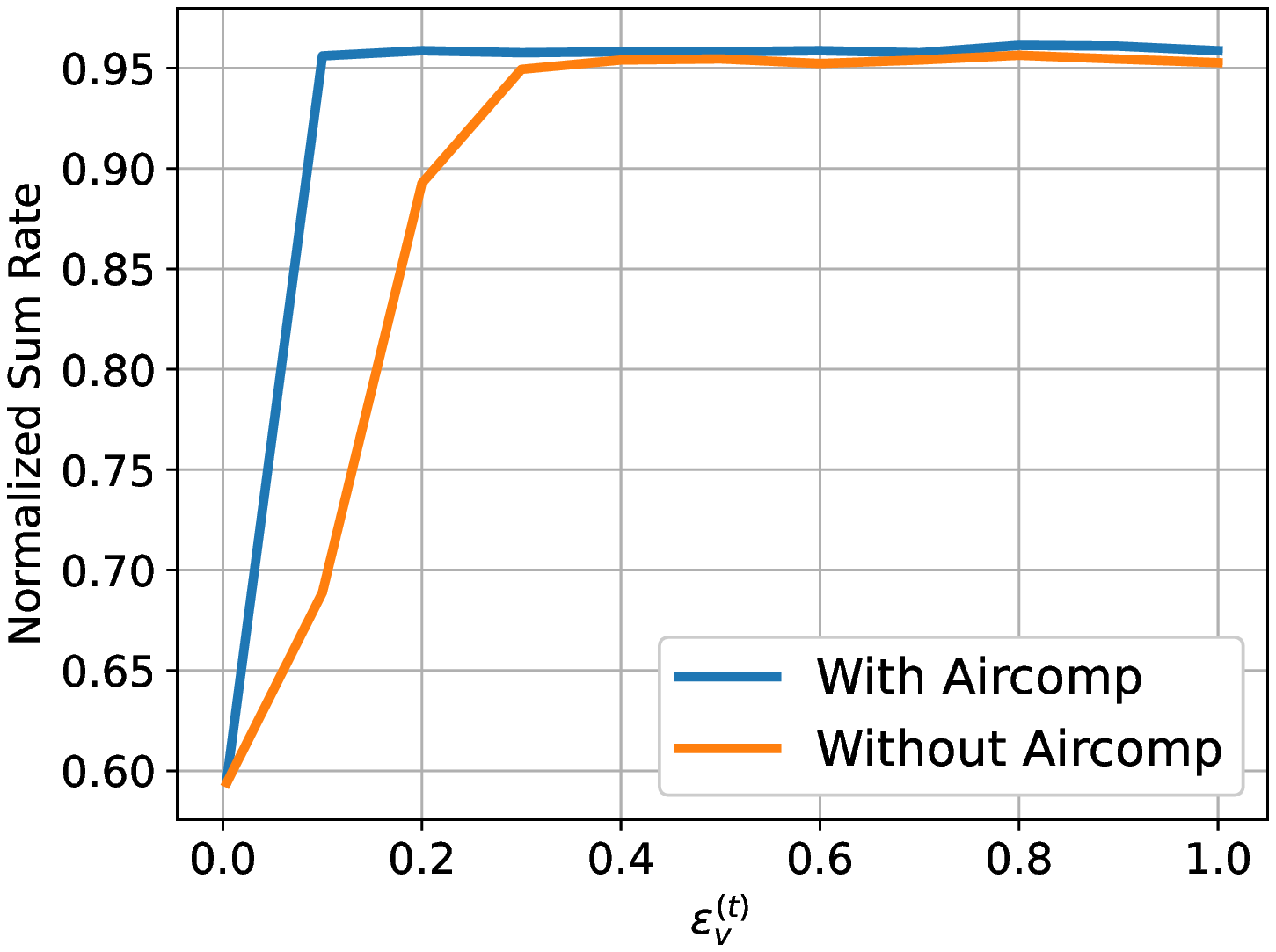}
			\label{fig:sumrate_comp}
		\end{minipage}
	}
	\vspace{-0.5em}
	\caption{Performance comparison between systems with and without Aircomp.}
	\label{fig:comp}
\end{figure}
In this part, we use simulation results to validate the advantage of the Aircomp technique and the simulation results are summarized in Fig. \ref{fig:comp}. As suggested by Fig. \ref{fig:comp}, the Aircomp technique can efficiently enhance the received SNR and the  normalized sum rate of each node especially under stringent privacy target, i.e., in the privacy-limited region. Therefore, the simulation results are consistent with and validate our theoretical analysis in Section \ref{section:comp_analysis}.

\section{Conclusion} \label{section:con}
Regarding decentralized control/management in the next-generation wireless communications, GNN is a potential technique since its inference stage can be naturally implemented in a decentralized manner. To achieve decentralized inference with GNNs, information exchanges among neighbors through wireless channels are inevitable, which however may lead to privacy leakage on personal features. To overcome this issue, this paper analyzes and enhances the privacy of decentralized inference with GNNs in wireless networks. We first design novel privacy-preserving wireless signaling for privacy-preserving decentralized inference with GNNs, based on which we analyze the SNR-privacy trade-off and the corresponding performance upper bound. Then the privacy-guaranteed training algorithm is proposed to further enhance decentralized inference performance with GNNs in wireless networks. Furthermore, we theoretically demonstrate that the Aircomp technique can simultaneously enhance the communication and computation efficiency as well as the privacy preservation. Extensive simulation results validate our theoretically analysis about the SNR-privacy trade-off and the performance gain brought by the Aircomp technique. The results also suggest that  our proposed privacy-preserving wireless signaling and privacy-guaranteed training algorithm work well for privacy-preserving decentralized inference with GNNs in wireless networks.

To the best of our knowledge, we are among the first to study privacy-preserving  decentralized inference with GNNs over noisy and fading wireless channels. As mentioned above, facilitating decentralized inference with GNNs is crucial for bringing GNN based wireless techniques from theory to practice. Unfortunately, related work is relatively rare in literature and some open problems still need further research endeavors. First, one can consider other techniques, such as the adaptive modulation and coding, the joint source-channel coding, or other LDP mechanisms to further enhance the privacy of decentralized inference with GNNs in wireless networks. Moreover, one can also pay attention to other obstacles for the practical implementation of GNNs in wireless communications, such as the synchronization among neighbors. Finally, this paper uses SNR to implicitly represent the performance of decentralized inference with GNNs in wireless networks. How to build a direct connection between the inference performance and the wireless environment is worth future investigation.

\begin{appendices}
\section{Proof of Theorem 1} \label{epsion1}
With the help of Definition 2, the achieved privacy guarantee by adopting the Gaussian mechanism depends on the $l_2$ sensitivity of the protected function and the variance of added Gaussian noise. By looking into (\ref{rv}), the variance of noise, including artificial and channel noise, is given by $\sigma_{vm}^2=\sum_{u\in N(v)}|g_{uv}|^2\beta_{uv}P_u+\sigma_v^2$. Furthermore, by considering two adjacent sub-graph data $G_v$ and $G'_v$ for node $v$ where only one neighbor feature differs, the $l_2$ sensitivity of the protected signal part is derived as follows

\begin{footnotesize}
\begin{equation*}
	\begin{aligned}
		\Delta_{2v} &= \max_{G_v, G'_v} ||\bm{\tilde{R}}_v^{(1)}(G_v)-\bm{\tilde{R}}_v^{(1)}(G_v')||_2 
		\\&= C_v \max_{\bm{h}_u^{(0)},\bm{h'}_u^{(0)}} ||\check{f}_M^{(1)}(\bm{h}_u^{(0)},\bm{e}_{vu})-\check{f}_M^{(1)}(\bm{h'}_u^{(0)},\bm{e}_{vu})||_2, \\
		&\leq  C_v \max_{\bm{h}_u^{(0)},\bm{h'}_u^{(0)}} ||\check{f}_M^{(1)}(\bm{h}_u^{(0)},\bm{e}_{vu})||_2+||\check{f}_M^{(1)}(\bm{h'}_u^{(0)},\bm{e}_{vu})||_2\overset{(a)}{=} 2C_v,
	\end{aligned}
\end{equation*}
\end{footnotesize}where step (a) depends on the fact that $||\check{f}_M^{(1)}(\bm{h}_u^{(0)},\bm{e}_{vu})||_2$ $ = 1$. In this way, for each node $u$ in $N(v)$,  $\bm{\tilde{R}}_v^{(1)}$ achieves $(\epsilon_v, \delta)$-LDP when predicting node $v$'s result, where
\begin{small}\begin{equation*}
	\begin{aligned}
		\epsilon_v &= \frac{\Delta_{2v}}{\sigma_{vm}}\sqrt{2\ln \frac{1.25}{\delta}}
		= \frac{2C_v}{\sqrt{\sum_{u\in N(v)}|g_{uv}|^2\beta_{uv}P_u+\sigma_v^2}}\sqrt{2\ln \frac{1.25}{\delta}}.
	\end{aligned}
\end{equation*}
\end{small}

\section{Proof of (\ref{rset1}) - (\ref{rset3})} \label{appendix:phi}
When $\epsilon_v^* \leq \epsilon_v^{(0)}$ holds,  $\rho_v^{\max} ={\epsilon_v^*}^2/(8\ln \frac{1.25}{\delta})$. The feasible solutions of $C_v$ and $\{\beta_{uv}\}_{u\in N(v)}$ fall into the set defined as 
\begin{small}
\begin{subnumcases}{}
	0< C_v \leq \min_{u\in N(v)}\sqrt{|g_{uv}|^2P_u}, & \label{set1} \\
	0 \leq \beta_{uv}\leq 1-\frac{C_v^2}{|g_{uv}|^2P_u}, \forall u\in N(v), &\label{set2} \\
	\frac{2C_v}{\sqrt{\sum_{u\in N(v)}|g_{uv}|^2\beta_{uv}P_u+\sigma_v^2}}\sqrt{2\ln \frac{1.25}{\delta}} = \epsilon_v^*.&\label{set3}
\end{subnumcases}
\end{small}By combining (\ref{set2}) and (\ref{set3}), we have
\begin{small}
\begin{equation*}
	\begin{aligned}
		\epsilon_v^*
		&\geq \frac{2C_v}{\sqrt{\sum_{u\in N(v)}|g_{uv}|^2(1-\frac{C_v^2}{|g_{uv}|^2P_u})P_u+\sigma_v^2}}\sqrt{2\ln \frac{1.25}{\delta}}
\end{aligned}
\end{equation*}
\begin{equation*}
	\begin{aligned}\rightarrow C_v \leq \epsilon_v^*\sqrt{\frac{\sigma_v^2+\sum_{u\in N(v)}|g_{uv}|^2P_u}{8\ln \frac{1.25}{\delta}+|N(v)|{\epsilon_v^*}^2}}.
	\end{aligned}
\end{equation*}
\end{small}Thus, the feasible solution set can be rewritten as (\ref{rset1}) - (\ref{rset3}).

\section{Proof of Theorem 2} \label{epsion2}
The proof is similar to that in Appendix \ref{epsion1}. By looking into (\ref{rv_no}), the variance of noise, including artificial and channel noise, is given by $\sigma_{vum}^2=|g_{uv}|^2\beta_{uv}P_u+\sigma_v^2$. Furthermore, the $l_2$ sensitivity of the protected signal part is similarly derived as (\ref{new2}).

\begin{figure*}[tb]
	\small
\begin{equation}
	\begin{aligned}
		\Delta_{2vu} &= \max_{\bm{h}_u^{(0)},\bm{h'}_u^{(0)}} || |g_{uv}|\sqrt{\alpha_{uv}P_u}\check{f}_M^{(1)}(\bm{h}_u^{(0)},\bm{e}_{vu})- |g_{uv}|\sqrt{\alpha_{uv}P_u}\check{f}_M^{(1)}(\bm{h'}_u^{(0)},\bm{e}_{vu})||_2, \\
		&=|g_{uv}|\sqrt{\alpha_{uv}P_u} \max_{\bm{h}_u^{(0)},\bm{h'}_u^{(0)}} ||\check{f}_M^{(1)}(\bm{h}_u^{(0)},\bm{e}_{vu})-\check{f}_M^{(1)}(\bm{h'}_u^{(0)},\bm{e}_{vu})||_2 \leq 2|g_{uv}|\sqrt{\alpha_{uv}P_u}.
  \end{aligned}
\label{new2}
\end{equation}
	\hrulefill
\end{figure*}
In this way, for node $u$ in $N(v)$,  $\bm{\tilde{R}}_{uv}^{(1)}$ achieves $(\epsilon_{uv}, \delta)$-LDP when predicting node $v$'s result, where

\begin{small}
\begin{equation*}
	\begin{aligned}
		\epsilon_{uv} &= \frac{\Delta_{2vu}}{\sigma_{vum}}\sqrt{2\ln \frac{1.25}{\delta}}
		= \frac{2|g_{uv}|\sqrt{\alpha_{uv}P_u}}{\sqrt{|g_{uv}|^2\beta_{uv}P_u+\sigma_v^2}}\sqrt{2\ln \frac{1.25}{\delta}}.
	\end{aligned}
\end{equation*}
\end{small}

\end{appendices}


\begin{thebibliography}{1}
	\bibitem{6g}
	W. Saad, M. Bennis, and M. Chen, ``A vision of 6G wireless systems: Applications, trends, technologies, and open research problems," \emph{IEEE Netw.}, vol. 34, no. 3, pp. 134--142, May/Jun. 2020.
	
	\bibitem{shi}
	H. Sun, X. Chen, Q. Shi, M. Hong, X. Fu, and N. D. Sidiropoulos, ``Learning to optimize: Training deep neural networks for interference management," \emph{IEEE Trans. Signal Process.}, vol. 66, no. 20, pp. 5438--5453, Oct. 2018.
	
	\bibitem{lorm}
	Y. Shen, Y. Shi, J. Zhang, and K. B. Letaief, ``LORM: Learning to optimize for resource management in wireless networks with few training samples," \emph{IEEE Trans. Wireless Commun.}, vol. 19, no. 1, pp. 665--679, Jan. 2020.
	
	\bibitem{learntobranch}
	M. Lee, G. Yu, and G. Y. Li, 	``Learning to branch: Accelerating resource allocation in wireless networks," \emph{IEEE Trans. Veh. Technol.}, vol. 69, no. 1, pp. 958--970, Jan. 2020.
	
	\bibitem{spatial}
	W. Cui, K. Shen, and W. Yu, ``Spatial deep learning for wireless scheduling," \emph{IEEE J. Sel. Areas Commun.}, vol. 37, no. 6, pp. 1248--1261, Jun. 2019.

    \bibitem{gnn_wireless_survey}
     M. Lee, G. Yu, H. Dai, and G. Y. Li, ``Graph neural networks meet wireless communications: Motivation, applications, and future directions,"  \emph{IEEE Wirel. Commun.}, vol. 29, no.5, pp. 12--19, Oct. 2022.
     
	 \bibitem{graphnn}
	M. Lee, G. Yu, and G. Y. Li, ``Graph embedding based wireless link scheduling with few training samples," \emph{IEEE Trans. Wireless Commun.},  vol. 20, no. 4, pp. 2282--2294, Apr. 2021.
	
	\bibitem{eisen_graph}
	M. Eisen and A. R. Ribeiro, ``Optimal wireless resource allocation with random edge graph neural networks," \emph{IEEE Trans. Signal Process.}, vol. 68, pp. 2977--2991, Apr. 2020.
	
	\bibitem{shen_graph}
	Y. Shen, Y. Shi, J. Zhang, and K. B. Letaief, ``Graph neural networks for scalable radio resource management: Architecture design and theoretical analysis," \emph{IEEE J. Sel. Areas Commun.}, vol. 39, no. 1, pp. 101--115, Jan. 2021.
	
	\bibitem{liu_graph}
	R. Liu,	M. Lee, G. Yu, and G. Y. Li, ``User association for millimeter-wave networks: A machine learning approach," \emph{IEEE Trans. Commun.}, vol. 67, no. 7, pp. 4162--4174, Jul. 2020.
	
	\bibitem{ca}
	M. Lee, S. Hosseinalipour, C. Brinton, G. Yu, and H. Dai, ``A fast graph neural network-based method for winner determination in multi-unit combinatorial auctions," \emph{IEEE Trans. Cloud Comput.}, vol. 10, no. 4, pp. 2264--2280,  Oct.-Dec. 2022.
	
	
	\bibitem{chen}
	X. Xu, Q. Chen, X. Mu, Y. Liu, and H. Jiang, ``Graph-embedded multi-agent learning for smart reconfigurable THz MIMO-NOMA networks," \emph{IEEE J. Sel. Areas Commun.}, vol. 40, no. 1, pp. 259--275, Jan. 2022.
	
	\bibitem{tmc}
	M. Lee, G. Yu, and H. Dai, ``Decentralized inference with graph neural networks in wireless communication systems,"  \emph{IEEE Trans. Mobile Comput.}, vol. 22, no. 5, pp. 2582–2598, May 2023.
	
	\bibitem{networking1}
	M. Wang, Y. Cui, X. Wang, S. Xiao, and J. Jiang, ``Machine learning for networking: WorkFlow, advances and opportunities," \emph{IEEE Netw.}, vol. 32, no. 2, pp. 92--99, Mar./Apr. 2018.
	
	\bibitem{networking2}
	E. Zeydan, E. Bastug, M. Bennis, M. A. Kader, I. A. Karatepe, A. S. Er, and M. Debbah, ``Big data caching for networking: Moving from cloud to edge," \emph{IEEE Commun. Mag.}, vol. 54, no. 9, pp. 36--42, Sep. 2016.
	
	\bibitem{physical1}
	T. O'Shea and J. Hoydis, ``An introduction to deep learning for the physical layer," \emph{IEEE Trans. Cogn. Commun. Netw.}, vol. 3, no. 4,	pp. 563--575, Dec. 2017.
	
	
	\bibitem{physical2}
	Z. Qin, H. Ye, G. Y. Li, and B.-H. F. Juang, ``Deep learning in physical layer communications," \emph{IEEE Wireless Commun.},  vol. 26, no. 2, pp. 93--99, Apr. 2019.
	
	\bibitem{gnn_channel_1}
      T. Jiang, H. V. Cheng, and W. Yu, ``Learning to reflect and to beamform for intelligent reflecting surface with implicit channel esti- mation," \emph{IEEE J. Sel. Areas Commun.}, vol. 39, no. 7, pp. 1931--1945, May 2021.

      \bibitem{gnn_channel_2}
      K. Tekbiyik, G. K. Kurt, C. Huang, A. R. Ekti, and H. Yanikomeroglu, ``Channel estimation for full-duplex RIS-assisted HAPS backhauling with graph attention networks," in \emph{Proc. IEEE ICC}, Jun. 2021, pp. 1--6.
 
      \bibitem{gnn_mimo}
      A. Scotti, N. N. Moghadam, D. Liu, K. Gafvert, and J. Huang, ``Graph neural networks for massive MIMO detection,"  \emph{arXiv preprint arXiv:2007.05703}, Jul. 2020.
	
	
	\bibitem{gnn_survey1}
	Z. Wu, S. Pan, F. Chen, G. Long, C. Zhang, and P. S. Yu, ``A comprehensive survey on graph neural networks," \emph{IEEE Trans. Neural Netw. Learn. Syst.}, vol. 32, no. 1, pp. 4--24, Jan. 2021.
	
	\bibitem{gnn_survey2}
	J. Zhou, G. Cui, S. Hu, Z. Zhang, C. Yang, Z. Liu, L. Wang, C. Li, and M. Sun,  ``Graph neural networks: A review of methods and applications," \emph{arXiv preprint arXiv:1812.08434}, Jul. 2019.
	
	\bibitem{control1}
 E. Tolstaya, F. Gama, J. Paulos, G. Pappas, V. Kumar, and A. Ribeiro, ``Learning decentralized controllers for robot swarms with graph neural networks," in \emph{Proc. Conf. Robot Learn.}, 2020, pp. 671--682.
	
\bibitem{control2}
Q. Li, F. Gama, A. Ribeiro, and A. Prorok, ``Graph neural networks for decentralized multi-robot path planning," in \emph{Proc. IEEE/RSJ Int. Conf.
Intell. Robots Syst.}, Oct. 2020, pp. 1--8.
	
\bibitem{control3}
F. Gama, E. Tolstaya, and A. Ribeiro, ``Graph neural networks for decentralized controllers,” in \emph{ICASSP 2021-2021 IEEE Int. Conf. on coust., Speech and Signal Process.}, Jun. 2021, pp. 5260--5264.

\bibitem{control4}
Z. Gao, Y. Shao, D. G\"{u}nd\"{u}z, and A. Prorok, ``Decentralized channel management in WLANs with graph neural networks," \emph{arXiv preprint arXiv:2210.16949
}, Oct. 2022.
	
\bibitem{fedprivacy}
C. Wu, F. Wu, Y. Cao, Y. Huang, and X. Xie, ``FedGNN: Federated graph neural network for privacy-preserving recommendation," \emph{arXiv preprint arXiv:2102.04925},  Mar. 2021.
	
\bibitem{splitprivacy}
C. Shan, H. Jiao, and J. Fu, ``Towards representation identical privacy-preserving graph neural network via split learning," \emph{arXiv preprint arXiv:2107.05917}, Jul. 2021.
	
\bibitem{private_sgd}
T. T. Mueller, J. C. Paetzold, C. Prabhakar, D. Usynin, D. Rueckert, and G.  Kaissis, ``Differentially private graph classification with GNNs," \emph{arXiv preprint arXiv:2202.02575}, Feb. 2022.

 
	\bibitem{private_adam}
	T. Igamberdiev and I. Habernal, ``Privacy-preserving graph convolutional networks for text classification," \emph{arXiv preprint arXiv:2102.09604}, Jul. 2021.
	
		\bibitem{private_noise}
	 A. Daigavane, G. Madan, A. Sinha, A. G. Thakurta, G. Aggarwal, and P. Jain, ``Node-level differentially private graph neural networks,"\emph{arXiv preprint arXiv:2111.15521}, Dec. 2021.
	
	\bibitem{private_coding}
	S. Sajadmanesh and D. Gatica-Perez, ``Locally private graph neural
	networks," in \emph{Proc. 2021 ACM SIGSAC}, Nov. 2021, pp. 2130--2145.
	
	 
	 \bibitem{wirelessprivate1}
	 M. Seif, R. Tandon, and M. Li, ``Wireless federated learning with local differential privacy," in \emph{2020 IEEE Int. Symp. Inf. Theory}, Jun. 2020, pp. 2604--2609.
	
	\bibitem{wirelessprivate2}
	 A. Sonee, S. Rini, and Y.-C. Huang, ``Wireless federated learning
	with limited communication and differential privacy," \emph{arXiv preprint arXiv:2106.00564}, Jun. 2021.

	\bibitem{wirelessprivate3}
	S. Chen, D. Yu, Y. Zou, J. Yu, and X. Cheng, ``Decentralized wireless federated learning with differential privacy,"  \emph{IEEE Trans. Industr. Inform.}, vol. 18, no. 9, pp. 6273--6282, Sep. 2022.

\bibitem{wirelessprivate4}
 D. Liu and O. Simeone, ``Privacy for free: Wireless federated learning via uncoded transmission with adaptive power control," \emph{IEEE J. Sel. Areas Commun.}, vol. 39, no. 1, pp. 170--185, 2021.

\bibitem{wirelessprivate5}
Y. Koda, K. Yamamoto, T. Nishio, and M. Morikura, ``Differentially
private aircomp federated learning with power adaptation harnessing receiver noise," \emph{arXiv preprint arXiv:2004.06337}, 2020.

\bibitem{wirelessprivate6}
K.Wei, J. Li, M. Ding, C. Ma, H. H. Yang, F. Farokhi, S. Jin, T. Q. Quek, and H. V. Poor, ``Federated learning with differential privacy: Algorithms and performance analysis," \emph{IEEE Trans. Inf. Forensics Secur.}, 2020.

\bibitem{wirelessprivate7}
A. Sonee and S. Rini, ``Efficient federated learning over multiple access channel with differential privacy constraints," emph{arXiv preprint arXiv:2005.07776}, 2020.

	\bibitem{google}
	\'{U}. Erlingsson, V. Pihur, and A. Korolova, ``RAPPOR: Randomized aggregatable privacy-preserving ordinal response," in \emph{Proc. ACM SIGSAC Conf. Comput. Commun. Secur.}, Scottsdale, AZ, USA, Nov. 2014,  pp. 1054--1067.
	
	\bibitem{Microsoft}
	B. Ding, J. Kulkarni, and S. Yekhanin, ``Collecting telemetry data
	privately," in \emph{Proc. Adv. Neural Inf. Process. Syst.}, 2017, pp. 3571--3580.
	
	\bibitem{dp}
	C. Dwork and A. Roth, ``The Algorithmic Foundations of Differential
	Privacy," \emph{Foundations and Trends$^\circledR$ in Theoretical Computer Science},
	vol. 9, no. 3-4, pp. 211--407, Aug. 2014.

 \bibitem{client_dp}
 Y. Zhao, J. Zhao, M. Yang, T. Wang, N. Wang, L. Lyu, D. Niyato, and
K.-Y. Lam, ``Local differential privacy-based federated learning for Internet of Things," \emph{IEEE Internet Things J.}, vol. 8, no. 11, pp. 8836--8853, Jun. 2021.
	
	\bibitem{FL_aircomp}
	G. Zhu, Y. Wang, and K. Huang, ``Low-latency broadband analog aggregation for federated edge learning," \emph{IEEE Trans. Wireless Commun.}, vol. 19, no. 1,	pp. 491--506, Jan. 2020.
	
	\bibitem{FL_aircomp2}
	 K. Yang, T. Jiang, Y. Shi, and Z. Ding, ``Federated learning via over-the-air computation," \emph{IEEE Trans. Wireless Commun.}, vol. 19, no. 3, pp. 2022--2035, Mar. 2020.
  
\bibitem{sync1}
  M. Goldenbaum and S. Stanczak, ``Robust analog function computation
via wireless multiple-access channels," \emph{IEEE Trans. Commun.},
vol. 61, no. 9, pp. 3863--3877, Aug. 2013.

\bibitem{sync2} 
O. Abari, H. Rahul, D. Katabi, and M. Pant, ``Airshare: Distributed
coherent transmission made seamless," in \emph{2015 IEEE INFOCOM}, Apr. 2015, pp. 1742--1750.
	
\bibitem{adam}
D. P. Kingma and J. Ba, ``Adam: A method for stochastic optimization," in \emph{Proc. 3rd Int. Conf. Learn. Represent. (ICLR)}, May 2014, pp. 1--6.
	
\bibitem{wmmse}
Q. Shi, M. Razaviyayn, Z.-Q. Luo, and C. He, ``An iteratively weighted MMSE approach to distributed sum-utility maximization for a MIMO interfering broadcast channel," \emph{IEEE Trans. Signal Process.}, vol. 59, no.9, pp. 4331--4340, Sept. 2011.
\end{thebibliography}
\end{document}